\documentclass[fleqn,12pt]{article}
\usepackage{graphicx}
\onecolumn
\newtheorem{tab}{Table}
\newtheorem{fig}{Figure}
\title{Phys.Teacher {\bf 48},4(2006) \\
 A SHORT HISTORY OF HINDU ASTRONOMY \& EPHEMERIS}
\author{P.Rudra\\ 
  Formerly of  Department of Physics, University of Kalyani, \\
   Kalyani, WB, 741-235, India\\  \\
   e-mail:~~p.rudra@saha.ac.in~~~prasanta.rudra@gmail.com\\  \\  \\
   PACS~no.  01.65.+g   95.10.Km  }
\date{ }
\begin{document}
\maketitle
\vskip 2.0cm
\samepage
\nopagebreak
\begin{abstract}
\indent {\hskip 0.4cm} We have summarized here the astronomical
knowledge of the ancient Hindu astronomers. This knowledge was
accumulated from before 1500 B.C. up to around 1200 A.D. In
Section \ref{equiv} we have correlated terms used by the Hindu
astronomers and their equivalents in modern astronomy. In Section 
\ref{coord} we have defined the different astronomical coordinate 
systems and their transformation relations.  In Sections
\ref{seasons} and \ref{cycles} we have collected the main features
of solar and lunar motions in terms of modern astronomical
terminology. In Section \ref{stars} we have given the names of
the stars mentioned by the Hindu astronomers and their modern
names with the present astronomical coordinates. In Section 
\ref{hist} we have given a short survey of Indian history with 
emphasis to Hindu astronomy. In Sections \ref{vedas} and \ref{sid} 
we have given short descriptions of the main sources of Hindu 
astronomy. In Section \ref{astron} the important features of Hindu 
astronomy have been described. All through the article we have tried 
to present the content in tabular forms for ready and unambiguous 
perception.
\end{abstract}
\pagebreak
\newpage
\section{Introduction \label{intro}}

\indent {\hskip 0.4cm} Every civilization from the antiquity to the present 
time has been fascinated by astronomy, study of motion of the heavenly
bodies across the sky. To-day astronomy has many tools of investigation:
$\gamma$-ray, $X$-ray, ultraviolet, visible, infra-red, microwave and
radio-wave radiations emanating from the heavenly bodies. This has
opened up a horizon that was not available to the ancients. They only
had the visible range of the spectrum. Their studies were thus
restricted to what we now call positional astronomy and did not
extend to investigations of the nature of these bodies. However,
extensive data about positional astronomy exist in ancient Egyptian, 
Babylonian, Indian and Chinese records.
\nopagebreak
\samepage

Here we have summarized the knowledge that ancient Hindu Astronomers
had arrived at. The time span is from circa 1500 B.C. to 1200 A.D.
This is a very long span of time. The technical terms used by these 
people have to be collated to present astronomical vocabularies.
These data were collected at places distributed over a vast region
of the earth. These facts complicate the process of evaluation of
correctness of these data.
\nopagebreak
\samepage

In Table \ref{equiv1} we have given the Sanskrit 
names of different astronomical terms and the corresponding technical 
names of present-day astronomy. In writing Sanskrit words in Roman
letters of alphabet we have followed the rule of transliteration
recommended by the International Alphabet of Sanskrit Transliteration
 (IAST) given in Table \ref{introtab}.
\nopagebreak
\samepage

  In Section \ref{coord} we 
explain the different coordinate systems used to describe the position 
of the heavenly bodies on the Celestial Sphere.  In Section \ref{seasons} 
we have described present-day knowledge of the motion of the sun and 
the moon, the two heavenly bodies most important to terrestrial beings.  
In Section \ref{cycles} we have mentioned the different lunar and
solar time periods that are of importance to astronomy. In Section
\ref{stars} we tabulate the fixed stars on the Celestial Sphere that 
were mentioned by Hindu Astronomers in their studies. In Section
\ref{hist} we describe the salient points of history of the Indians,
their interaction with Western Asia. A map (Figure \ref{mapfig})
gives us the idea of the region that is of importance to the
development of Hindu Astronomy. The Vedas and the Siddh\={a}ntas
are the main texts that tell us about the range of knowledge of the
Hindu astronomers. In Sections \ref{vedas} and \ref{sid} we 
describe the salient features of these sources. In Section \ref{astron}
the principal features of Hindu astronomy over the ages are
summarized.  
\pagebreak
\newpage
\begin{tab}
{\rm Rule of transliteration from Sanskrit to Roman letters of
alphabet according to International Alphabet of Sanskrit 
Transliteration (IAST)}
 \label{introtab}
\begin{center}
\includegraphics[scale=0.80]{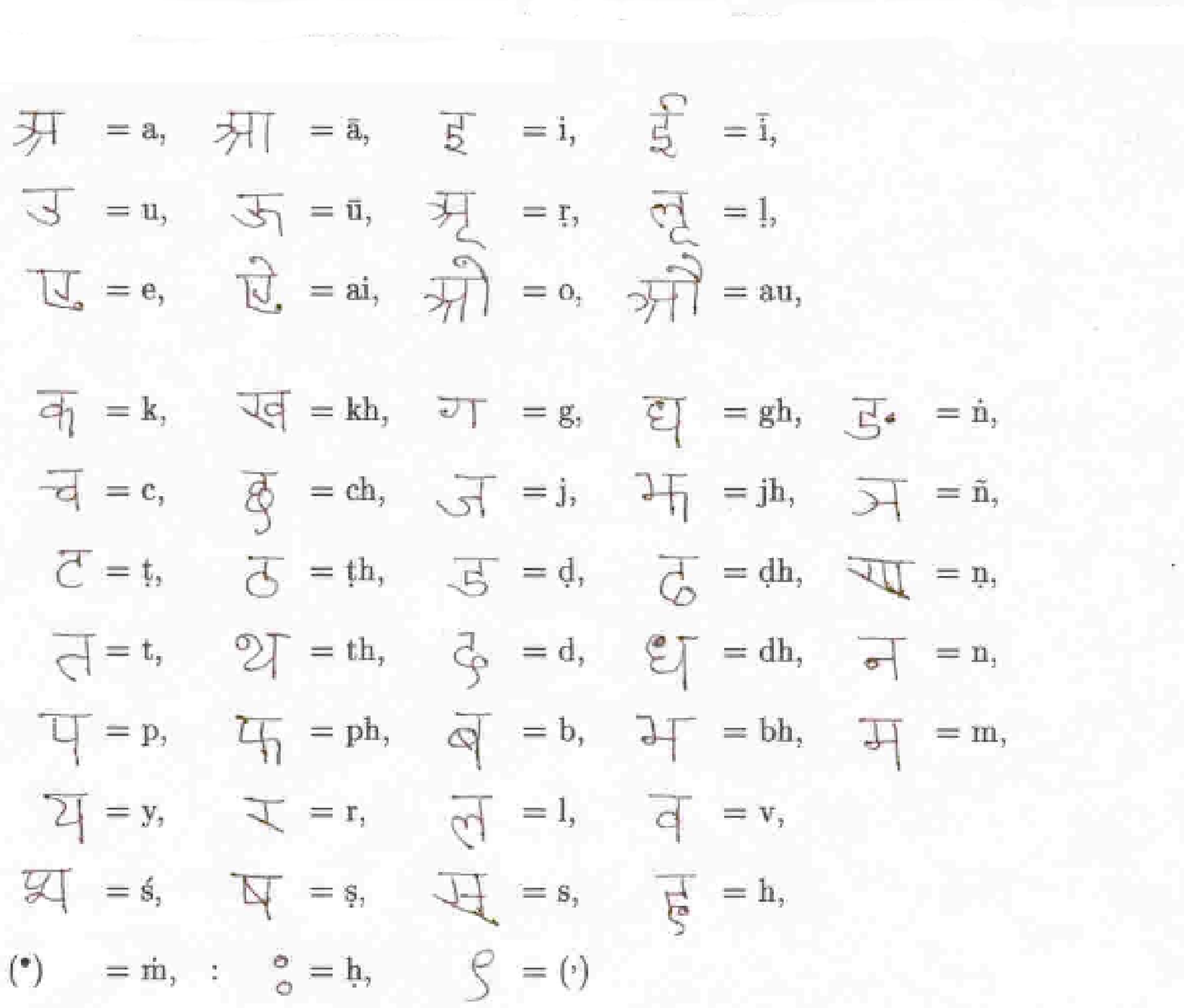}
\end{center}
\end{tab}

\section{Equivalence of Astronomical Terms  \label{equiv}}

\indent {\hskip 0.4cm} In all records of Hindu Astronomy we find mentions
of different astronomical terms. These are given in Table \ref{equiv1}
along with their modern equivalents.  \\
\begin{tab}   
{\rm Technical terms in Modern Astronomy and their equivalents in
 Hindu Astronomy.  \label{equiv1}}
\end{tab}
\begin{tabular}{||c|c||} \hline  \hline
Modern Astronomical Term & Equivalent Term in Hindu Astronomy \\  \hline \hline
Altitude & Madhyonnati \\ \hline
Apogee & Apabh\={u} \\ \hline
Apse & Apad\={u}raka \\ \hline
Aries, first point of & Me\d{s}\={a}di \\ \hline
Astronomy & Jyoti\d{s}a \\ \hline
Azimuth & Dig\={a}\.{m}\'{s}a \\ \hline
Declination & Vi\d{s}uvalamba \\ \hline
Ecliptic & Kr\={a}ntiv\d{r}tta \\ \hline
Equator, Celestial & Mah\={a}vi\d{s}uva \\ \hline
Equinox, Autumnal & Jalavi\d{s}uva Sa\.{m}kr\={a}nti \\ \hline
Equinox, Vernal & Mah\={a}vi\d{s}uva Sa\.{m}kr\={a}nti \\ \hline
Horizon plane & Digcakra \\ \hline
Hour Angle & Hor\={a}ko\d{n}a \\ \hline
Libra, first point of & Tul\={a}di \\ \hline
Luminary & Jyoti\d{s}ka \\ \hline
Meridian & Madhyarekh\={a} \\ \hline
Moon, Full & P\={u}r\d{n}im\={a} \\ \hline
Moon, New & Am\={a}basy\={a} \\ \hline
Perigee & Anubh\={u} \\ \hline
Polaris & Dhruvat\={a}r\={a} \\ \hline
Right Ascension & Vi\d{s}uv\={a}\.{m}\'{s}a \\ \hline
Saros & Yuga \\ \hline
Sidereal & N\={a}k\d{s}atra \\ \hline
Solstice, Summer & Karka\d{t}a Sa\.{m}kr\={a}nti \\ \hline
Solstice, Winter & Makara Sa\.{m}kr\={a}nti \\ \hline
Sphere, Celestial & Khagola \\  \hline
Zenith & Khamadhya \\ \hline
Zodiac signs & R\={a}\'{s}icakra \\ \hline  \hline
\end{tabular}

If we go through this table we find that the Hindu Astronomers had 
known the basic facts of positional astronomy regarding motion of
the heavenly bodies through the Celestial Sphere, the different
planes of motion of the sun and the moon, and the zodiac signs of
the fixed stars.

\section{Coordinates describing an object {\it X} \label{coord}}

\indent {\hskip 0.4cm} In positional astronomy, motion of a heavenly
body $X$ is described in different coordinate systems. In ordinary
3-dimensional space with spherical symmetry, spherical polar
coordinate system is the natural choice. On the Celestial Sphere, the
real distance from the observer is not of any importance. Only the
two angular coordinates fix the position of a body on the Celestial
Sphere. This requires an azimuthal plane and a reference direction
on this plane. For the coordinate of a place on the surface of the
earth (Cartographic system) we take the terrestrial equatorial plane
as the azimuthal plane and Greenwich's meridian supplies the
starting position for measuring the azimuthal angle. The topocentric
(Horizon) system is suitable for direct measurement by an observer
on earth's surface. For this the local horizon plane supplies the
azimuthal plane. The geocentric (Equatorial) system, which is
independent of the observer's position on earth's surface, uses the
Celestial Equator (the plane of intersection of the terrestrial
equator with the Celestial Sphere) as the azimuthal plane. Description
in terms of the geocentric coordinate system frees the observational
data from the dependence on the position of the local observer on
earth's surface. Figure \ref{geofig} shows the topocentric and the
geocentric coordinate systems. To describe the motion of the planets
and the stars on the Celestial Sphere, rotational and non-uniform
yearly motion of earth's motion around the sun has to be avoided.
The heliocentric (Ecliptic) system takes Ecliptic (the plane of sun's
motion on the Celestial Sphere) as the azimuthal plane. Figure 
\ref{heliofig} shows the heliocentric coordinate system. In Table
\ref{coordtab} we have compared the 3 astronomical coordinate systems
and the terrestrial cartographic system. The laws of transformation
from one system to another are as follows: \\
\vskip 0.5cm
$\begin{array}{lll} \left( \begin{array}{c} \cos \delta \cos \chi \\
 -\cos \delta \sin \chi \\ \sin \delta \end{array} \right) & = &
 \left( \begin{array}{ccc} \cos \bar{\phi} & 0 & \sin \bar{\phi} \\
 0 & 1 & 0 \\ -\sin \bar{\phi} & 0 & \cos \bar{\phi} \end{array}
 \right)~\left( \begin{array}{c} \cos \zeta \cos \psi \\
 -\cos \zeta \sin \psi \\ \sin \zeta \end{array} \right)~,
 \end{array}$ \\
\vskip 0.5cm
$\begin{array}{lll} \left( \begin{array}{c} \cos \beta \cos \lambda \\
 \cos \beta \sin \lambda \\ \sin \beta \end{array} \right) & = &
 \left( \begin{array}{ccc} 1 & 0 & 0 \\ 0 & \cos {\epsilon}_0 &
 \sin {\epsilon}_0 \\ 0 & -\sin {\epsilon}_0 & \cos {\epsilon}_0
 \end{array} \right)~\left( \begin{array}{c} \cos \delta \cos \alpha \\
 \cos \delta \sin \alpha \\ \sin \delta \end{array} \right)
 \end{array}$ \\
\vskip 0.5cm
{\sl Azimuth\/} ${\psi}_0$ of a Star at setting ($\zeta ~=~0$) is given by
$\sin ({\psi}_0~-~90^{\circ})~=~\sin \delta / \cos \phi$. \\
For the Sun at {\sl Summer Solstice~(S.S.)\/}, 
$\delta~=~23^{\circ}~30^{\prime}$.
\pagebreak
\newpage
\begin{fig} 
{\rm Topocentric and Geocentric Coordinates.}
 \label{geofig}
\begin{center}
\includegraphics[scale=0.50]{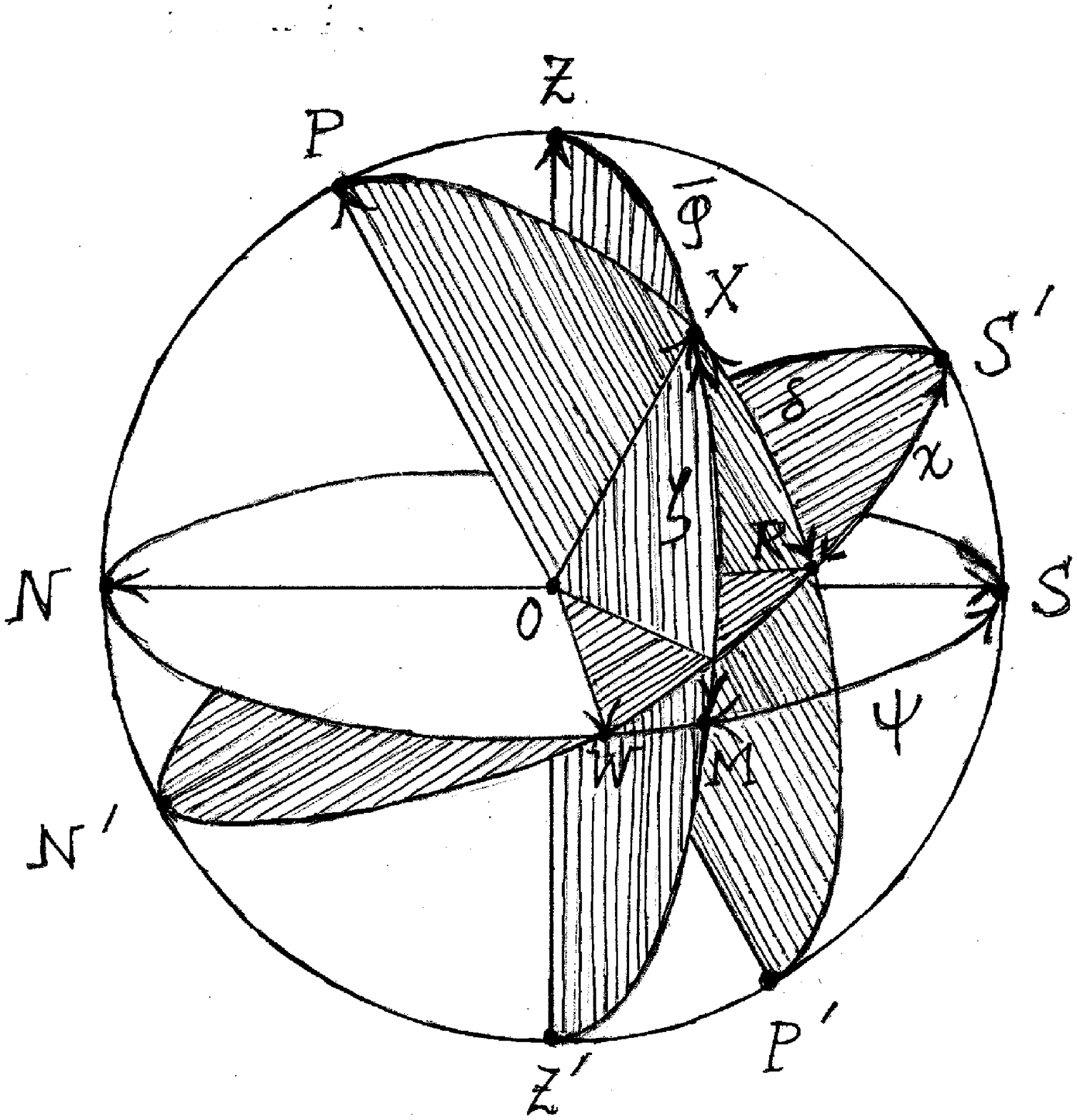}
\end{center}
\end{fig}
$\begin{array}{llllll} O & = & {\rm Observer,} & X & = &
 {\rm Astronomical~Object} \end{array}$  \\
$\begin{array}{lllllllll} Z & = & Zenith, &   & SWNE & \equiv &
 {\rm Plane}~\perp ~OZ & = & Horizon~Plane,\\
 &   & (Khamadhya)  &   &   &   &   &   &   (Digcakra) \\
 P & = &  Polaris, &   & S^{\prime}WN^{\prime}E & \equiv &
 {\rm Plane}~\perp ~OP & = & Celestial~Equator,\\
   &   & (Dhruvat\bar{a}r\bar{a}) &   &   &   &   &   & (Mah\bar{a}vi\d{s}uva)
 \end{array}$ \\
$\begin{array}{lll} {\rm Observer's}~Meridian & = & {\rm Great~Circle~through~}
  P~{\rm and~}Z \\ {\rm Object's}~Meridian & = & {\rm Great~Circle~through~}
  P~{\rm and~}X \\  \end{array}$ \\
$\begin{array}{lll}
 \left. {\begin{array}{lll} N & \equiv & North \\ S & \equiv & South 
 \end{array}} \right\} & = & \left\{ \begin{array}{l}{\rm Points~of~ 
 intersection~of~the} \\ {\rm Observer's}~Meridian~{\rm with~the~}
 Horizon~Plane, \end{array} \right. \\
 \left. \begin{array}{lll} E & \equiv & East  \\ W & \equiv & West 
 \end{array} \right\} & = & \left\{ \begin{array}{l} {\rm Points~of~
 intersection~of~the} \\ Celestial~Equator~{\rm with~the~} Horizon~Plane, 
 \end{array} \right.  \end{array}$ \\
$\begin{array}{lll} N^{\prime},~S^{\prime} & = & {\rm Points~of~ 
 intersection~of~the} \\  &   & {\rm Observer's}~Meridian~{\rm with~the~}
 Celestial~Equator, \end{array}$  \\
$\begin{array}{lllllllllll} \zeta & = & MX & \equiv & Altitude~ 
 ({\it Madhyonnati}), &   & \delta & = & RX & \equiv & Declination~ 
 ({\it Vi\d{s}uvalamba}) \\ \psi & = & SM & \equiv & Azimuth~ 
 ({\it Dig\bar{a}\dot{m}\acute{s}a}), &   & \chi & = & S^{\prime}R & \equiv &
 Hour~Angle \end{array}$ \\
$\begin{array}{lllll} Sidereal~time & \equiv & \tau & = &
 Hour~Angle~{\rm of~a~fixed~feature~}Q~{\rm on~the~}Celestial~Sphere 
  \end{array}$  \\
$\begin{array}{lllll} Right~Ascension~(R.A.)~{\rm of~}X & \equiv &
 \alpha & = & \tau~-~\chi    \end{array}$ \\
$\alpha$ increases towards $E$ from $Q$, opposite to the apparent
motion of the {\sl Celestial Sphere}
\pagebreak
\newpage
\begin{fig}  
{\rm Heliocentric Coordinates.}
   \label{heliofig}
\begin{center}
\includegraphics[scale=0.50]{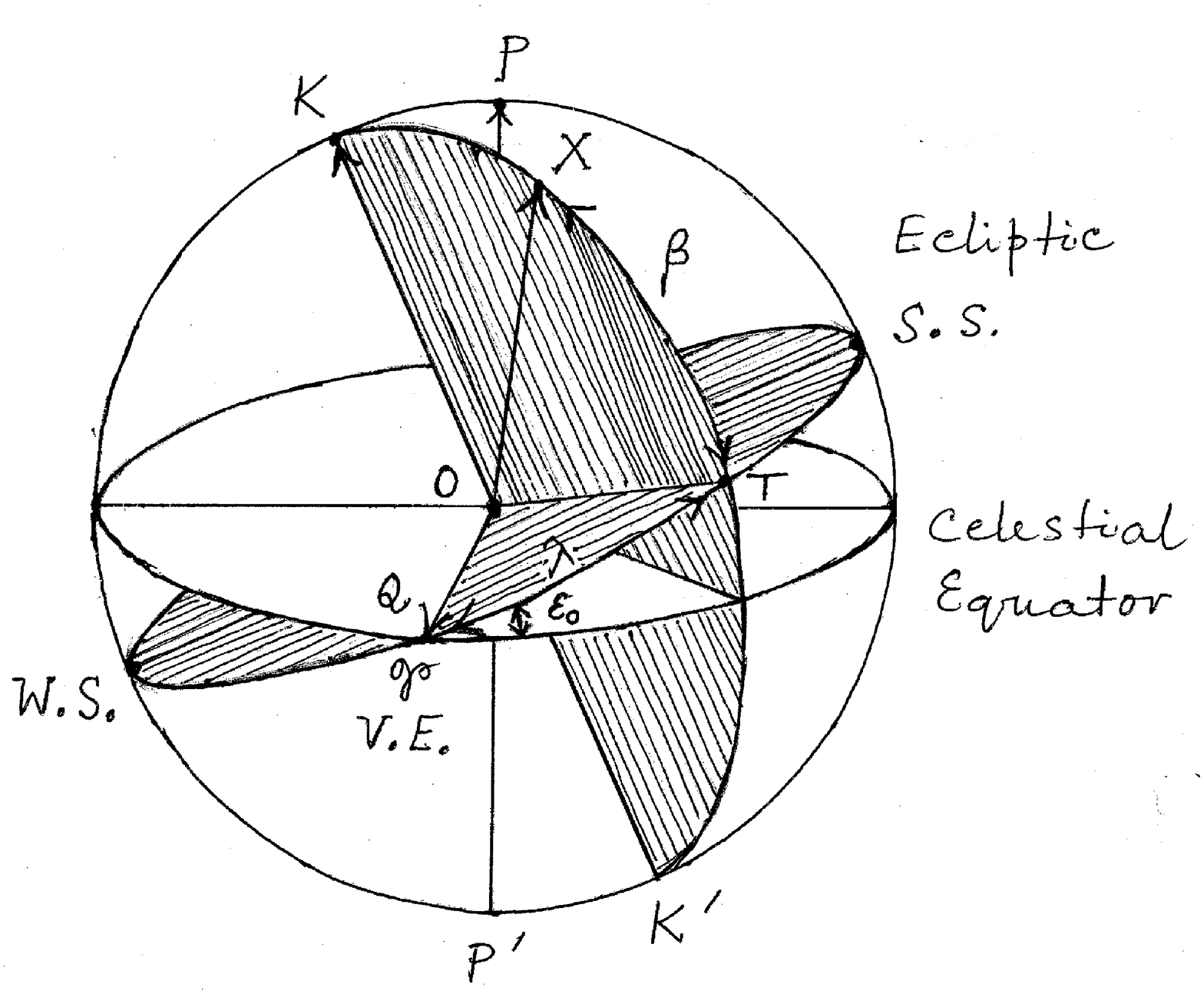}
\end{center}
\end{fig}
$\begin{array}{lll} Ecliptic & = & {\rm Plane~of~the~motion~of~the~Sun~
 on~the~}Celestial~Sphere, \\
 K & = & {\rm Pole~of~the~}Ecliptic~{\rm on~the~}Celestial~Sphere, \\
 {\epsilon}_0 & \equiv & {\rm obliquity~of~the~}Ecliptic \\
  &   & ~~~{\rm with~respect~to~the~}Celestial~Equator \\
  & = & \left\{\begin{array}{l} 23^{\circ}~30^{\prime}~~(\rm in~1950~A.D.) \\
 23^{\circ}~51^{\prime}~(in~1300~B.C.) \end{array} \right. \end{array}$ \\
$\begin{array}{lll} \Upsilon ~(~V.E.)~{\rm and~}\Omega ~(~A.E.) & = &
 {\rm Points~of~intersection~of~the~}Ecliptic \\
 &   &  {\rm and~the~}Celestial~Equator   \end{array}$ \\

$\Upsilon$ has an involved retrograde motion with respect to the fixed stars.
Its {\sl Celestial Longitude\/} with respect to a given fixed star decreases
by about $50^{\prime \prime}$ per year. \\
In antiquity $\Upsilon$ was in the {\sl Constellation of Aries\/} 
({\sl Me\d{s}a\/}), but now $\Upsilon$ has moved to the {\sl Constellation
of Pisces\/} ({\sl M\={i}na\/}). \\
The fixed feature {\sl Q\/} is taken to be the {\sl Vernal~Equinox~} point, 
$\Upsilon $.
\vskip 0.3cm 
$\begin{array}{lllll}  \beta & = & TX & = & Celestial~Latitude, \\   
 \lambda & = & \Upsilon~T & = & Celestial~Longitude \end{array}$  \\
\begin{tab} 
{\rm Coordinates used in different Astronomical Syatems. }
 \label{coordtab}
\end{tab}
\begin{tabular}{||c||c||c||c||} \hline  \hline
{\bf Cartographic} & {\bf Topocentric} & {\bf Geocentric} &
 {\bf Heliocentric}    \\ 
 System & {Horizon system} & {Equatorial system} & {Ecliptical system}
 \\  \hline   \hline
$\begin{array}{c} \phi \\ =~{\it Geographical} \\ {\it Latitude,} \\
 {\rm measured} \\ {\rm from~the}\\ {\it Terrestrial} \\ {\it Equator} \\
 {\rm along~the} \\ {\it Object's} \\ {\it Meridian,} \\
 \bar{\phi}~=~90^{\circ} - \phi \\ =~{\it Colatitude} \end{array}$ &
$\begin{array}{c} \zeta ~=~{\it MX} \\ =~{\it Altitude} \\ ({\it Madhyonnati}),
 \\ {\rm measured} \\ {\rm from~the} \\ {\it Horizon~plane} \\
 {\rm towards~the} \\ Zenith  \end{array}$ &
$\begin{array}{c} \delta ~=~{\it RX} \\ =~{\it Declination}\\ 
 ({\it Vi\d{s}uvalamba\/}), \\ {\rm measured} \\ {\rm from~the} \\
 {\it Celestial} \\ {\it Equator} \\ {\rm along~the} \\ {\it Object's} \\
 {\it Meridian}  \end{array}$ & 
$\begin{array}{c} \beta ~=~{\it TX} \\ =~{\it Celestial} \\ {\it Latitude,}\\
 {\rm measured} \\ {\rm from~the} \\ {\it Ecliptic} \\ {\rm towards~} K 
 \end{array}$ \\  \hline
$\begin{array}{c} {\it Geographical} \\ {\it Longitude,} \\
 {\rm measured~from} \\ {\it Greenwich's} \\ {\it Meridian} \\
 {\rm along~the} \\ {\it Terrestrial} \\ {\it Equator} \end{array}$ &
$\begin{array}{c} \psi ~=~{\it SM} \\ =~{\it Azimuth} \\
 ({\it Dig\bar{a}\dot{m}\acute{s}a\/}), \\ {\rm measured} \\
 {\rm from}~{\it South} \\ {\rm towards}~{\it West} \\ {\rm along~the} \\
 {\it Horizon~plane} \end{array}$ &
$\begin{array}{c} \chi ~=~{\it S}^{\prime}{\it R} \\
  =~{\it Hour~Angle,} \\ {\rm measured} \\ {\rm from~the}\\
 {\it Observer's} \\{\it Meridian} \\ {\rm towards}~{\it West} \\
 {\rm along~the} \\ {\it Celestial} \\ {\it Equator} \end{array}$ &
$\begin{array}{c} \lambda~=~\Upsilon {\it T} \\ =~{\it Celestial} \\
 {\it Longitude,} \\ {\rm measured~from}\\ \Upsilon~{\rm towards~the} \\
 {\it Summer~Solstice} \\ {\rm along~the} \\ {\it Ecliptic} \end{array}$ \\
  \hline  \hline
\end{tabular}  \\

\section{Sun ({\sl Sol}) - Earth ({\sl Terra}) - Moon ({\sl Luna}) System
   \label{seasons}} 

\indent {\hskip 0.4cm} To people living on the surface of the earth,
the sun and the moon are the two most important heavenly bodies. In
Figure \ref{terrafig} we show our present-day knowledge of the motion of the 
earth around the sun throughout the year. This heliocentric picture was, 
however, not available to the people of antiquity. What they observed was 
the Ecliptic, the plane of solar motion on the Celestial Sphere, and the
difference of speed of the sun along the Ecliptic at different seasons
of the year. \\

\begin{fig} 
{\rm Earth's motion around the Sun.}
  \label{terrafig} 
\begin{center}
\includegraphics[scale=0.60]{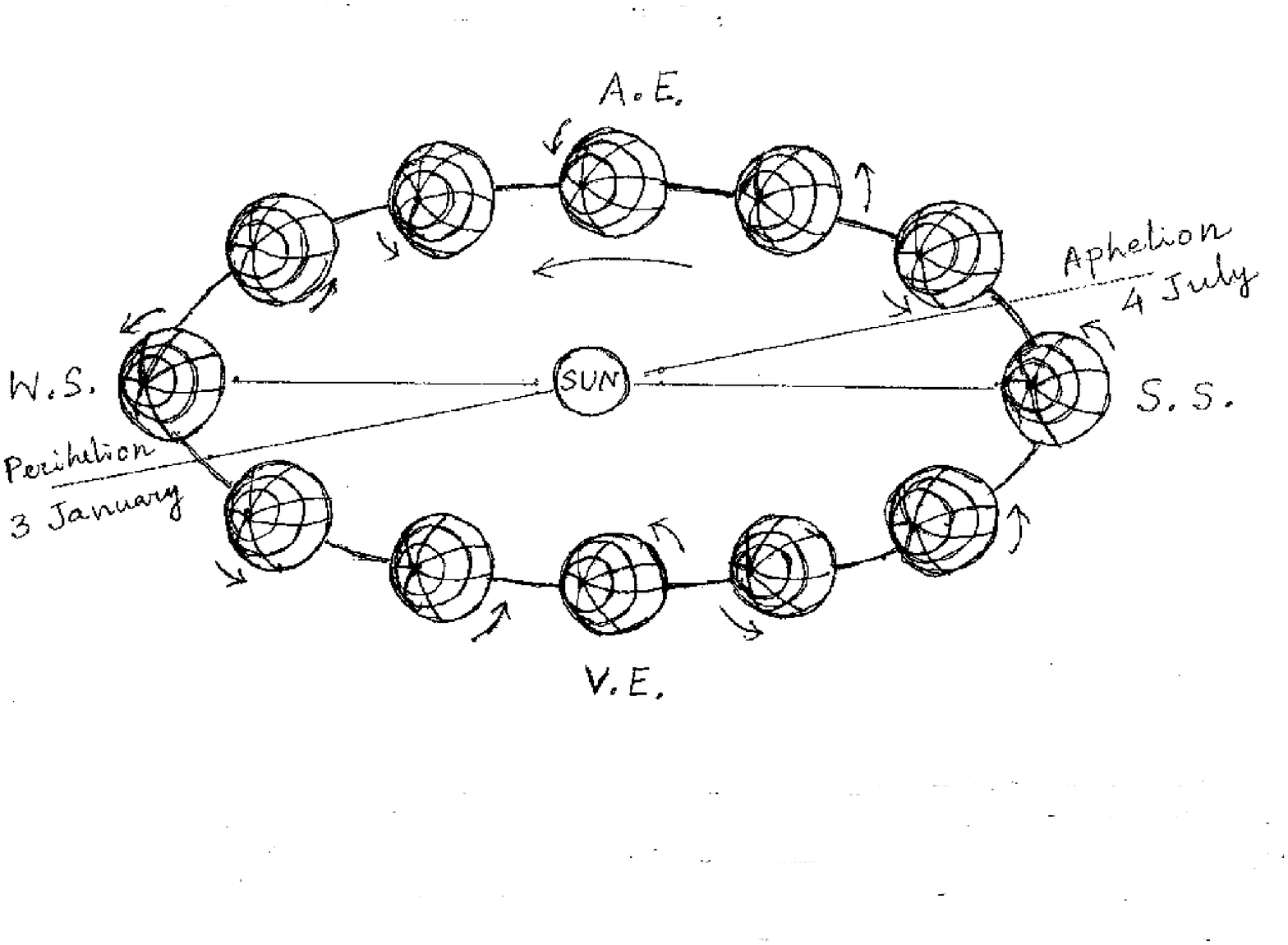}
\end{center}
\end{fig} 
$\begin{array}{lllll} V.E. & \equiv & Vernal~Equinox~
 (Mah\bar{a}vi\d{s}uva~Sa\dot{m}kr\bar{a}nti) & = & 21~{\rm March~in~1950~A.D.} 
 \\  A.E. & \equiv & Autumnal~Equinox~(Jalavi\d{s}uva~Sa\dot{m}kr\bar{a}nti) &
 = & 22~{\rm September~in~1950~A.D.} \\
 W.S. & \equiv & Winter~Solstice~(Makara~Sa\dot{m}kr\bar{a}nti) & = &
 22~{\rm December~in~1950~A.D.} \\
 S.S. & \equiv & Summer~Solstice~(Karka\d{t}a~Sa\dot{m}kr\bar{a}nti) & = &
 21~{\rm June~in~1950~A.D.} \\
 \end{array}$ \\

Exact motion of Earth's axis has a number of periodic terms called
{\sl nutations\/} $\sim ~9^{\prime \prime}$.
\indent {\hskip 0.2cm} Similarly lunar motion around the earth is
shown in Figure \ref{lunafig}. This motion, of course, tallied with 
observations from earth's surface. 
\begin{fig} 
{\rm Moon's motion around the Earth.}
  \label{lunafig} 
\begin{center}
\includegraphics[scale=0.70]{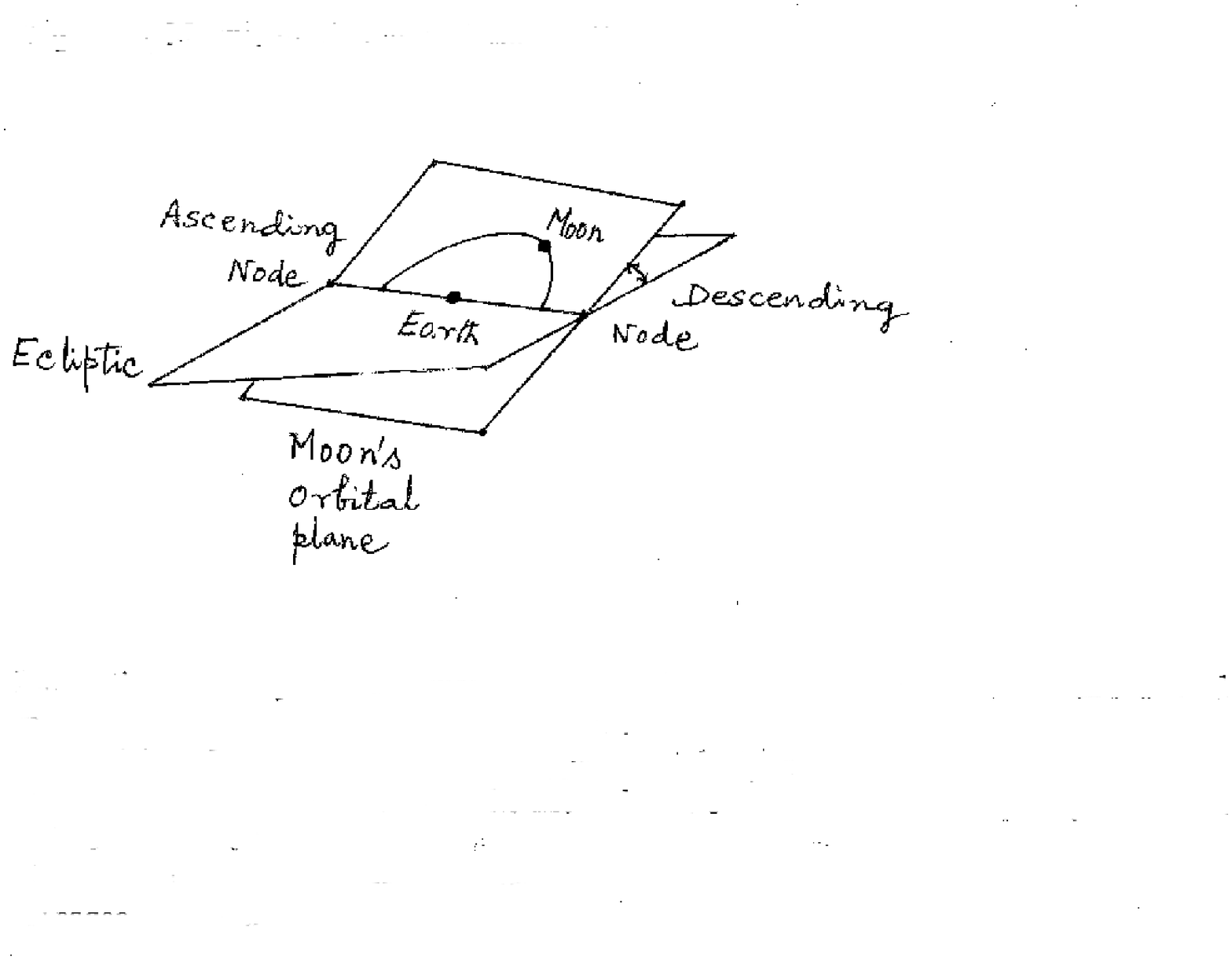}
\end{center}
\end{fig}
Inclination of the plane of Lunar motion to the {\sl Ecliptic} 
(the plane of Earth's annual motion around the Sun) is approximately 
$5^{\circ}$.  \\
Eccentricity of Lunar elliptic orbit is approximately 1/18.  \\
Upper limit of the declination of the Moon has a cycle of
approximately 18 years (from $18^{\circ}~10^{\prime}$ to
$28^{\circ}~50^{\prime}$). \\
$\begin{array}{lll} Perigee & = & {\rm Point~of~closest~approach,} \\
 Apogee & = & {\rm Point~of~farthest~approach} \end{array}$

\section{Lunar and Solar Cycles  \label{cycles}}

\indent {\hskip 0.4cm} The two most important cyclic motions are the 
lunar and the solar cycles. They provide us with the two most easily
observable time periods. Since actual periods are not exactly
constant all through the year due to non-uniform motion of the earth
around the sun, a number of different associated periods have been
used by the astronomers. These are given in Table \ref{timetab}.

{\sl X transits\/} or {\sl culminates\/} when the {\sl Meridian\/} of
{\sl X\/} coincides with the {\sl Meridian\/} of the Observer. \\
$\begin{array}{lll} Sidereal~Day~{\rm of~} X & = & {\rm from~one~}
 transit~{\rm of~} X~ {\rm to~the~next~} transit, \\
 Mean~Solar~Day~(M.S.D.) & = & Sidereal~Day~ {\rm of~a~fictitious~} 
 Mean~Sun~ (M.S.~) \\ &   & {\rm moving~with~a~uniform~speed} \\
 &   & {\rm on~the~} Celestial~Equator, \\  
 Equation~of~time & = & {\alpha}_{M.S.}~-~{\alpha}_S,  \\
 Greenwich~Mean &   &   \\ Astronomical~time~(GMAT) & = &
 Hour~Angle~{\rm of~} M.S., \\
 Universal~time~(UT) & = & GMAT~+~12~hrs, \\
 {\rm Mean~midnight} & = & 0~hr, \\
 {\rm Mean~noon} & = & 12~hr  \end{array}$ \\
\begin{tab}  
{\rm Different Lunar and Solar Time Periods.}
  \label{timetab}
\end{tab}
\begin{tabular}{||l||l|l|l||}  \hline  \hline
 Motion & Name (Symbol) & Description & Present value in {\sl M.S.D.} \\
 \hline  \hline
  & {\sl Tropical Month,\/} $T_1$ & {\sl Equinox\/} to {\sl Equinox\/} &
 27.32158 \\  \cline{2-4} 
  & {\sl Anomalistic Month,\/} $T_2$ & {\sl Perigee\/} to {\sl Perigee\/} &
 27.53455 \\  \cline{2-4} 
 {\sl Lunar\/} & $\begin{array}{l} Draconic~(Nodical) \\ Month,~T_3
 \end{array}$ & {\sl Node\/} to {\sl Node\/} & 27.21222 \\  \cline {2-4}
  & {\sl Sidereal Month\/} & $\begin{array}{l} {\rm Return~to~the} \\
 {\rm same~fixed~Star} \end{array}$ & 27.32166 \\ \cline{2-4}
  & {\sl Synodic Month,\/} $T$ & $\begin{array}{l} New~Moon~{\rm to} \\
 New~Moon \\ (Am\bar{a}nta) \end{array}$ & 29.53059 \\
 \hline  \hline
 {\sl Solar\/} & $\begin{array}{l} Tropical~Year,~T_0 \\ (S\bar{a}va\d{n}a~
 Vatsara) \end{array}$ & {\sl Equinox\/} to {\sl Equinox\/} & 365.2422 \\
 \cline{2-4}  
  & {\sl Sidereal Year\/} & $\begin{array}{l} {\rm Return~to~the} \\
 {\rm same~fixed~Star} \end{array}$ & 365.257 \\ \hline  \hline
\end{tabular}

Hindu astronomers determined transits of heavenly bodies by 
{\sl Gnomon} (Sun-dial) and time was measured by {\sl Clepsydra} 
(Water-clock). Names of different subdivisions of time given in Table
\ref{subtime} indicate the possibility that Hindu astronomers were
able to measure these short time intervals. In fact in all treatises 
of Hindu Astronomy accurate directions for constructing water-clocks
are given \cite{diksit}.  \\
\begin{tab} 
{\rm Different subdivisions of time in Hindu Astronomy.}
  \label{subtime}
\end{tab}
\begin{tabular}{lllllllllll} 
1 Anupala & = & $\frac{1}{150}$ sec, &   &   &   &   &   &   &   &    \\
60 Anupala & = & 1 Bipala & = & 0.4 sec, &   & 60 Bipala & = & 1 Pala & = &
 24 sec, \\
60 Pala & = & 1 Da\d{n}\d{d}a & = & 24 min, &   & 2 Da\d{n}\d{d}a & = &
 1 Muh\={u}rta & = & 48 min  \\
7.5 Da\d{n}\d{d}a & = & 1 Prahara, & = & 3 hr, &    & 8 Prahara & = & 
 1 Divasa &
 = & 24 hr
\end{tabular}
\vskip 1.0cm
The unit of Muh\={u}rta has a significance in astronomy. It is the time
interval by which the time of moonrise is delayed on successive days. 

\section{Stars and their positions  \label{stars}}

\indent {\hskip 0.4cm} All astromical and ephemerical calculations are
based on motion of the heavenly bodies with respect to the fixed stars
on the Celestial Sphere. Hindu Astronomers also used the same
convention. The Celestial Sphere is mentally divided in 27 divisions
and each division is assigned to the most important star there. Out
of these 27 stars 12 are selected to give the names of the months.
Present-day Celestial Longitudes and Latitudes of these stars are
given in Table \ref{startab}. Modern Astronomical names of these
stars are also given. \\
\begin{tab} 
{\rm Names of {\sl Nak\d{s}atra\/}s in {\sl Ved\={a}\.{n}ga Jyoti\d{s}a,
S\={u}rya Siddh\={a}nta}, corresponding names in {\sl Modern Astronomy} and 
their {\sl Celestial Longitudes} and {\sl Latitudes} in 1950 A.D. The names of
the Lunar months are according to the stars written in bold face letters.}
   \label{startab}
\end{tab}
\begin{tabular}{||c||c|c||c||c|c||}   \hline  \hline
{Serial} & \multicolumn{2}{c||}{Name of {\sl Nak\d{s}atra\/} in} & 
 {Principal} & \multicolumn{2}{c||}{\sl Celestial}  \\ 
  \cline{2-3}   \cline{5-6}
{Number} & {\sl Ved\={a}\.{n}ga\/} & {\sl S\={u}rya\/} & {Star} &
 {\sl Longitude\/} & {\sl Latitude} \\ &  {\sl Jyoti\d{s}a\/} & 
 {\sl Siddh\={a}nta\/}  &    &    &    \\   \hline  \hline
1. & {K\d{r}ttik\={a}} &  {\bf K\d{r}ttik\={a}} & $\eta$ - Tauri &
 $~59^{\circ}~17^{\prime}~39^{\prime \prime}$ &
 $+~~4^{\circ}~~2^{\prime}~46^{\prime \prime}$ \\  \hline
2. & {Rohi\d{n}\={i}} &  {Rohi\d{n}\={i}} & $\alpha$ - Tauri &
 $~69^{\circ}~~5^{\prime}~25^{\prime \prime}$ &
 $-~~~5^{\circ}~28^{\prime}~14^{\prime \prime}$ \\  \hline
3. & {M\d{r}ga\'{s}\={i}r\d{s}a} &  {\bf M\d{r}ga\'{s}\={i}r\d{s}a} &
 $\lambda$ - Orionis & $~83^{\circ}~~0^{\prime}~31^{\prime \prime}$ &
 $-~~13^{\circ}~22^{\prime}~32^{\prime \prime}$ \\  \hline
4. & {\={A}rdr\={a}} &  {\={A}rdr\={a}} & $\alpha$ - Orionis &
 $~88^{\circ}~~3^{\prime}~22^{\prime \prime}$ &
 $-~~16^{\circ}~~1^{\prime}~59^{\prime \prime}$ \\  \hline
5. & {Punarvasu} &  {Punarvasu} & $\beta$ - Geminorum &
 $112^{\circ}~31^{\prime}~29^{\prime \prime}$ &
 $+~~~6^{\circ}~40^{\prime}~51^{\prime \prime}$ \\  \hline
6. & {Ti\d{s}ya} &  {\bf Pu\d{s}y\={a}} & $\delta$ - Cancri  &
 $128^{\circ}~~1^{\prime}~23^{\prime \prime}$ &
 $+~~~0^{\circ}~~4^{\prime}~32^{\prime \prime}$ \\  \hline
7. & {\={A}\'{s}re\d{s}\={a}} &  {A\'{s}le\d{s}\={a}} & $\epsilon$ - Hydrae  &
 $131^{\circ}~38^{\prime}~59^{\prime \prime}$ &
 $-~~11^{\circ}~~6^{\prime}~25^{\prime \prime}$ \\  \hline
8. & {Magh\={a}} &  {\bf Magh\={a}} & $\alpha$ - Leonis  &
 $149^{\circ}~~8^{\prime}~~1^{\prime \prime}$ &
 $+~~~0^{\circ}~27^{\prime}~48^{\prime \prime}$ \\  \hline
9. & $\begin{array}{c} {\rm P\bar{u}rva} \\ {\rm Phalguni} \end{array}$ &
 $\begin{array}{c} {\rm P\bar{u}rva} \\ {\rm Phalguni} \end{array}$ &
 $\delta$ - Leonis  & $160^{\circ}~36^{\prime}~52^{\prime \prime}$ &
 $+~~14^{\circ}~19^{\prime}~58^{\prime \prime}$ \\  \hline
10. & $\begin{array}{c} {\rm Uttara} \\ {\rm Phalguni} \end{array}$ &
 $\begin{array}{c} {\rm Uttara} \\ {\bf Phalguni} \end{array}$ &
 $\beta$ - Leonis  & $170^{\circ}~55^{\prime}~23^{\prime \prime}$ &
 $+~~12^{\circ}~16^{\prime}~13^{\prime \prime}$ \\  \hline
11. & {Hasta} &  {Hasta} & $\delta$ - Corvi  &
 $192^{\circ}~45^{\prime}~23^{\prime \prime}$ &
 $-~~12^{\circ}~11^{\prime}~31^{\prime \prime}$ \\  \hline
12. & {Citr\={a}} &  {\bf Citr\={a}} & $\alpha$ - Virginis  &
 $203^{\circ}~~8^{\prime}~37^{\prime \prime}$ &
 $-~~~2^{\circ}~~3^{\prime}~~4^{\prime \prime}$ \\  \hline
13. & {Sv\={a}t\={i}} &  {Sv\={a}t\={i}} & $\alpha$ - Bootis  &
 $203^{\circ}~32^{\prime}~~8^{\prime \prime}$ &
 $+~~30^{\circ}~46^{\prime}~~3^{\prime \prime}$ \\  \hline
14. & {Vi\'{s}\={a}kh\={a}} &  {\bf Vi\'{s}\={a}kh\={a}} & $\alpha$ - Librae  &
 $224^{\circ}~23^{\prime}~~7^{\prime \prime}$ &
 $+~~~0^{\circ}~20^{\prime}~19^{\prime \prime}$ \\  \hline  \hline
\end{tabular}
\pagebreak
\newpage
Table \ref{startab} (Continued)  \\
\begin{tabular}{||c||c|c||c||c|c||}   \hline  \hline
{Serial} & \multicolumn{2}{c||}{Name of {\sl Nak\d{s}atra\/} in} & 
 {Principal} & \multicolumn{2}{c||}{\sl Celestial}  \\ 
  \cline{2-3}   \cline{5-6}
{Number} & {\sl Ved\={a}\.{n}ga\/} & {\sl S\={u}rya\/} & {Star} &
 {\sl Longitude\/} & {\sl Latitude} \\ &  {\sl Jyoti\d{s}a\/} & 
 {\sl Siddh\={a}nta\/}  &    &    &    \\   \hline  \hline
15. & {Anur\={a}dh\={a}} & {Anur\={a}dh\={a}} & $\delta$ - Scorpii  &
 $241^{\circ}~52^{\prime}~23^{\prime \prime}$ &
 $-~~~1^{\circ}~58^{\prime}~49^{\prime \prime}$ \\  \hline
16. & $\begin{array}{c} {\rm Rohi\d{n}\bar{i},} \\ {\rm Jye\d{s}\d{t}h\bar{a}}
 \end{array}$ & $\begin{array}{c} {\bf Jye\d{s}\d{t}h\bar{a}} \end{array}$ &
 $\alpha$ - Scorpii  & $249^{\circ}~ 3^{\prime}~51^{\prime \prime}$ &
 $-~~33^{\circ}~33^{\prime}~50^{\prime \prime}$ \\  \hline
17. & $\begin{array}{c} {\rm Vic\d{r}tau,}\\ {\rm M\bar{u}labarha\d{n}\bar{i}}
 \end{array}$ & $\begin{array}{c} {\rm M\bar{u}l\bar{a}} \end{array}$ &
 $\lambda$ - Scorpii  & $263^{\circ}~53^{\prime}~14^{\prime \prime}$ &
 $-~~13^{\circ}~46^{\prime}~56^{\prime \prime}$ \\  \hline
18. & {P\={u}rv\={a}\d{s}\={a}\d{d}h\={a}} & 
 {\bf P\={u}rv\={a}\d{s}\={a}\d{d}h\={a}} & $\delta$ - Sagittarii  &
 $273^{\circ}~52^{\prime}~55^{\prime \prime}$ &
 $-~~~6^{\circ}~27^{\prime}~58^{\prime \prime}$ \\  \hline
19. & {Uttar\={a}\d{s}\={a}\d{d}h\={a}} & 
 {Uttar\={a}\d{s}\={a}\d{d}h\={a}} & $\sigma$ - Sagittarii  &
 $281^{\circ}~41^{\prime}~11^{\prime \prime}$ &
 $-~~~3^{\circ}~26^{\prime}~36^{\prime \prime}$ \\  \hline
  & {Abhijit} & {Abhijit} & $\alpha$ - Lyrae  &
 $284^{\circ}~36^{\prime}~54^{\prime \prime}$ &
 $+~~61^{\circ}~44^{\prime}~~7^{\prime \prime}$ \\  \hline
20. & {\'{S}ro\d{n}\={a}} & {\bf \'{S}rava\d{n}\={a}} & $\alpha$ - Aquilae  &
 $301^{\circ}~~4^{\prime}~16^{\prime \prime}$ &
 $+~~29^{\circ}~18^{\prime}~~18^{\prime \prime}$ \\  \hline
21. & {\'{S}ravi\d{s}\d{t}h\={a}} & {Dhani\d{s}\d{t}h\={a}} & 
 $\beta$ - Delphini  & $315^{\circ}~38^{\prime}~38^{\prime \prime}$ &
 $+~~31^{\circ}~55^{\prime}~~21^{\prime \prime}$ \\  \hline
22. & {\'{S}atabhi\d{s}ak} & {\'{S}atabhi\d{s}ak} & $\lambda$ - Aquarii  &
 $340^{\circ}~52^{\prime}~38^{\prime \prime}$ &
 $-~~~0^{\circ}~23^{\prime}~~8^{\prime \prime}$ \\  \hline
23. & $\begin{array}{c} {\rm P\bar{u}rva} \\ {\rm Pro\d{s}\d{t}hapada}
 \end{array}$ & $\begin{array}{c} {\rm P\bar{u}rva} \\ 
 {\rm Bh\bar{a}drapada} \end{array}$ & $\alpha$ - Pegasi  &
 $352^{\circ}~47^{\prime}~19^{\prime \prime}$ &
 $+~~19^{\circ}~24^{\prime}~25^{\prime \prime}$ \\  \hline
24. & $\begin{array}{c} {\rm Uttara} \\ {\rm Pro\d{s}\d{t}hapada}
 \end{array}$ & $\begin{array}{c} {\rm Uttara} \\ 
 {\rm Bh\bar{a}drapada} \end{array}$ & $\gamma$ - Pegasi  &
 $~~8^{\circ}~27^{\prime}~32^{\prime \prime}$ &
 $+~~12^{\circ}~35^{\prime}~55^{\prime \prime}$ \\  \hline
25. & {Revat\={i}} & {Revat\={i}} & $\zeta$ - Piscium  &
 $~19^{\circ}~10^{\prime}~40^{\prime \prime}$ &
 $-~~~0^{\circ}~12^{\prime}~52^{\prime \prime}$ \\  \hline
26. & {A\'{s}vayuja} & {\bf A\'{s}vin\={i}} & $\beta$ - Arietis  &
 $~33^{\circ}~16^{\prime}~18^{\prime \prime}$ &
 $+~~~8^{\circ}~29^{\prime}~~7^{\prime \prime}$ \\  \hline
27. & {Apabhara\d{n}\={i}} & {Bhara\d{n}\={i}} & $41$ - Arietis  &
 $~47^{\circ}~30^{\prime}~19^{\prime \prime}$ &
 $+~~10^{\circ}~26^{\prime}~48^{\prime \prime}$ \\  \hline
\end{tabular}

\section{Summary of Indian History  \label{hist}}

\indent {\hskip 0.4cm} Hindu Astronomy developed through about two
thousand years and this development took place not at a fixed region.
Parallel development of Astronomy also took place at Babylon and
Alexandria. Figure \ref{mapfig} shows the part of the world and
cities having importance in this historical development. Positions of
modern cities near these ancient seats of astronomy are also shown.  

A short summary of the History of India and
her interaction with Western Asia is given in Table \ref{histtab}.
The earliest time in Indian History that is of importance from
astronomical interest is the time of \d{R}k-vedas. This is the
oldest of all the Vedas (Section \ref{vedas}) and astronomical references 
found here fixes its time (terminus ad quem) around 1500 B.C., when
the Middle Kingdom of Egypt ended. The part of the Vedas known as
Ved\={a}\.{n}ga was compiled around 1200 B.C. when Egypt had the
New Kingdom. Time of Gautama Buddha (end of the 6th century B.C.)
is a fixed point in Indian History. Written history is available
afterwards. Mah\={a}bh\={a}rata, the great Indian epic was perhaps
compiled around 400 B.C. Alexander's invasion of the Northwestern
periphery of India in 323 B.C. and Candragupta's victory over
Seleucus in 306 B.C. brought India in contact with Babylonian
astronomy through the Greeks. But there is scarcely any evidence of
influence of Egyptian or Babylonian astronomy on Hindu astronomy at
this time. Advent of the \'{S}akas (Scythians) in Balkh (Bactria)
and their victory over the P\={a}radas (Parthians) in 123 B.C. would
later have profound impact on Hindu astronomy and ephemeris. The
Old \'{S}aka Era started from this momentous event. With the beginning of 
the reign of Ka\d{n}i\d{s}ka, the Ku\d{s}\={a}\d{n}a king, in 78 A.D.  
(201 in the Old \'{S}aka Era) the New \'{S}aka Era (omitting the 200 of
the Old \'{S}aka Era) began in India. This is the Era {\it par
excellence\/} in Hindu astronomy and no other Era could take its place
in Hindu ephemeris. The \'{S}akas grafted Babylonian system on
Hindu astronomy and by 400 A.D. the Siddh\={a}nta astronomy
(Section \ref{sid}) replaced the older Ved\={a}\.{n}ga astronomy.
\={A}ryabha\d{t}a, perhaps the greatest Hindu mathematician cum
astronomer, was born at Kusumapura near P\={a}\d{t}aliputra
(modern Patna) in 476 A.D. and his monumental work \={A}ryabha\d{t}iya
was compiled in 499 A.D. The next important event (6th century A.D.)
of Hindu astronomy is Var\={a}hamihira's Pa\~{n}ca-siddh\={a}ntik\={a},
second only to the appearence of \={A}ryabha\d{t}iya. The last
important contribution to Hindu astronomy came from Bh\={a}skar\={a}c\={a}rya
of Devagiri in 1207. After Muhammad Ghouri's conquest of Northern
India at the end of the 12th century A.D. all the academic centres 
of excellence in Northern India were completely destroyed and starved
of financial support. This saw the end of ancient Hindu astronomy.
\pagebreak
\newpage
\begin{fig}  
{\rm India and her Western neighbors.}
  \label{mapfig}
\begin{center}
\includegraphics[scale=0.90]{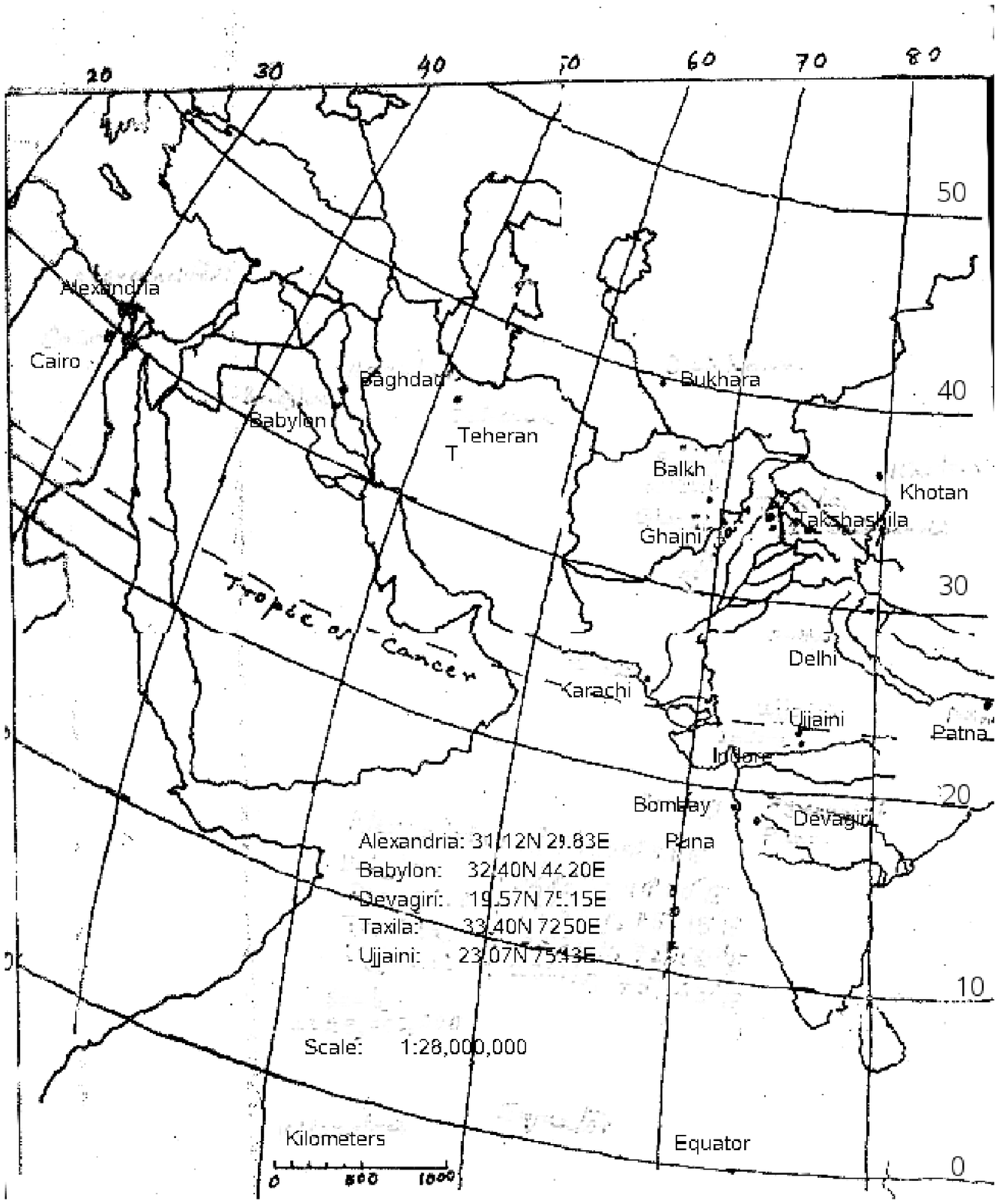}
\end{center}
\end{fig}
\pagebreak
\newpage
\begin{tab} 
{\rm History of India and her Western neighbors.}
 \label{histtab}
\end{tab}
\begin{tabular}{||l|l||l|l||}  \hline  \hline
{\sl Era\/} & {\sl Century\/} & {\sl Indian\/} & 
 {\sl West Asian\/}  \\  \hline  \hline
 & 13-th & $\begin{array}{l} \bullet~1207~{\rm Siddh\bar{a}nta~\acute{s}iro}
 {\rm ma\d{n}i} \\ {\rm ~~~of~}{\bf Bh\bar{a}skar\bar{a}c\bar{a}rya} \\
 {\rm ~~~of~Devagiri,} \\ {\rm ~~~(last~Indian} \\ {\rm ~~~contributor} \\
 {\rm ~~~of~importance} \end{array}$ & \\  \cline{2-4}
 & 12-th & $\begin{array}{l} \bullet~1192~{\rm Muhammad~Ghouri} \\
 {\rm ~~~defeats~P\d{r}thv\bar{i}r\bar{a}ja} \\ {\rm ~~~in~the~2nd~battle} \\
 {\rm ~~~of~Tarain} \end{array}$  &    \\  \cline{2-4}
 & 11-th &   &    \\  \cline{2-4}
A.D. & 10-th & $\begin{array}{l} \bullet~1000~{\rm Mahmud~of} \\
 {\rm ~~~Ghazni~invades} \\ {\rm ~~~India} \\
 \bullet~966~{\bf Bha\d{t}\d{t}ap\bar{a}la's} \\ {\rm ~~~Ga\d{n}itaskandha} \\
 \bullet~950~{\bf \bar{A}ryabha\d{t}a~II} \end{array}$  &  \\ \cline{2-4}
 & 9-th &   &    \\  \cline{2-4}
 & 8-th &   &    \\  \cline{2-4}
 & 7-th & $\begin{array}{l} \bullet~630~{\rm Har\d{s}avardhana} \\
 \bullet~628~{\bf Brahmagupta} \\ {\rm ~~~of~Khandakh\bar{a}dya} \\
  {\rm ~~~\&~Brahmasiddh\bar{a}nta} \end{array}$ &
 $\begin{array}{l} \bullet~700~{\rm Beginning} \\ {\rm ~~~of~Arab} \\
 {\rm ~~~civilization} \\ {\rm ~~~of~Baghdad,} \\ {\rm ~~~Damascus,} \\
 {\rm ~~~Cordova} \end{array}$  \\  \cline{2-4}
 & 6-th & $\begin{array}{l} \bullet~587~{\rm Death~of} \\ 
 {\bf ~~~Var\bar{a}hamihira~}{\rm of} \\
 {\rm ~~~Pa\tilde{n}ca~siddh\bar{a}ntik\bar{a}}
 \\ \bullet~580~{\bf \acute{S}r\bar{i}\d{s}e\d{n}a} \\
 \bullet~523~{\rm Death~of} \\ {\bf ~~~\bar{A}ryabha\d{t}a} \end{array}$ &  
 $\begin{array}{l} \bullet~600~{\rm End~of} \\ {\rm ~~~Sassanid} \\
 {\rm ~~~empire} \\ {\rm ~~~in~Iran} \end{array}$ \\  \cline{2-4}
 & $\cdots$ & $\cdots$ & $\cdots$ \\  \hline \hline
\end{tabular}
\pagebreak
\newpage 
Table \ref{histtab} (Continued) \\
\begin{tabular}{||l|l||l|l||}  \hline  \hline
{\sl Era\/} & {\sl Century\/} & {\sl Indian\/} & 
 {\sl West Asian\/}  \\  \hline  \hline
 & $\cdots$ & $\cdots$ & $\cdots$ \\   \cline{2-4}
 & 5-th & $\begin{array}{l} \bullet~500~{\bf Vi\d{s}\d{n}ucandra} \\
 \bullet~499~{\rm Composition} \\ {\rm ~~~of~\bar{A}ryabha\d{t}iya} \\
 \bullet~476~{\rm Birth~of} \\ {\bf ~~~\bar{A}ryabha\d{t}a} \\
 {\rm ~~~at~P\bar{a}\d{t}aliputra} \end{array}$  &   \\  \cline{2-4}
 & 4-th & $\begin{array}{l} \bullet~400~{\rm Beginning~of} \\
 {\rm ~~~Siddh\bar{a}nta~Jyoti\d{s}a} \\ \bullet~320~{\rm Beginning~of} \\
 {\rm ~~~Gupta~Empire} \end{array}$  &  \\  \cline{2-4}
A.D. & 3-rd &   & $\begin{array}{l} \bullet~224~{\rm Beginning~of} \\
 {\rm ~~~Sassanid~Empire} \\ {\rm ~~~in~Iran} \end{array}$ \\  \cline{2-4}
 & 2-nd & $\begin{array}{l} \bullet~170~{\rm \acute{S}akas~reign} \\
 {\rm ~~~at~Ujjain\bar{i}} \end{array}$  & 
 $\begin{array}{l} \bullet~150~{\bf Ptolemy} \\ {\rm ~~~of~Alexandria}
 \end{array}$   \\  \cline{2-4}
 & 1-st & $\begin{array}{l} \bullet~100~{\rm S\bar{a}tab\bar{a}hana} \\
 {\rm ~~~empire~ends} \\ \bullet~78~{\rm New~\acute{S}aka~Era,} \\
 {\rm ~~~Ka\d{n}i\d{s}ka's~reign} \\ {\rm ~~~begins} \\
 \bullet~50~{\rm K\bar{a}\d{n}va~dynasty} \\ {\rm ~~~at~P\bar{a}\d{t}aliputra,}
 \\ {\rm ~~~\acute{S}akas~rule~at} \\ {\rm ~~~Puru\d{s}apura} \\
 {\rm ~~~\&~Mathur\bar{a}} \end{array}$ & 
 $\begin{array}{l} \bullet~100~{\rm Selucid~rule~at} \\ {\rm ~~~Babylon~ends}
 \end{array}$  \\ \hline
 & 1-st & $\begin{array}{l} \bullet70~{\rm \acute{S}akas~capture} \\
 {\rm ~~~Tak\d{s}a\acute{s}il\bar{a}} \\ \bullet~85~{\rm Sunga~reign} \\
 {\rm ~~~ends} \end{array}$ & $\begin{array}{l} \bullet~80~{\rm \acute{S}akas~}
 {\rm occupy} \\ {\rm ~~~Afghanistan} \end{array}$ \\  \cline{2-4}
B.C. & 2-nd & $\begin{array}{l} \bullet~175~{\rm Menander} \\
 {\rm ~~~captures~Punjab~\&~Sind} \\ \bullet~180~{\rm Sunga~dynasty} \\
 {\rm ~~~at~P\bar{a}\d{t}aliputra,} \\ {\rm ~~~Selucids~capture} \\
 {\rm ~~~Tak\d{s}a\acute{s}il\bar{a}} \end{array}$ & 
 $\begin{array}{l} \bullet~123~{\rm \acute{S}akas~(Scythians)} \\
 {\rm ~~~defeat~P\bar{a}radas~(Parthians)} \\ {\rm ~~~\&~occupy~Balkh~} 
 {\rm (Bactria),}\\ {\rm ~~~Old~\acute{S}aka~era~begins} \\
 \bullet~138~{\rm End~of~Parthian}\\ {\rm ~~~empire~in~Iran} \end{array}$ \\
  \cline{2-4}
 & $\cdots$ & $\cdots$ & $\cdots$ \\  \hline \hline
\end{tabular}
\pagebreak
\newpage 
Table \ref{histtab} (Continued) \\
\begin{tabular}{||l|l||l|l||}  \hline  \hline
{\sl Era\/} & {\sl Century\/} & {\sl Indian\/} & 
 {\sl West Asian\/}  \\  \hline  \hline
 & $\cdots$ & $\cdots$ & $\cdots$ \\  \cline{2-4} 
 & 3-rd & $\begin{array}{l} \bullet~220~{\rm Beginning~of} \\
 {\rm ~~~\acute{S}\bar{a}tav\bar{a}hana} \\ {\rm ~~~empire} \\
 \bullet~226~{\rm Death~of~A\acute{s}oka} \\
 \bullet~268~{\rm A\acute{s}oka's~reign~begins} \end{array}$   & 
 $\begin{array}{l} \bullet~248~{\rm Beginning~of} \\ {\rm ~~~Parthian~Empire} \\
 {\rm ~~~in~Iran} \\ \bullet~290~{\bf Archimedes} \\ \bullet~300~{\bf Euclid,}\\
 {\rm ~~~End~of~Achaemenian} \\ {\rm ~~~Empire~in~Iran}
 \end{array}$ \\  \cline{2-4}
 & 4-th & $\begin{array}{l} \bullet~306~{\rm Candragupta} \\ {\rm ~~~defeats~}
 {\rm Seleucus} \\ \bullet~323~{\rm Alexander~invades} \\ {\rm ~~~India} \\
 \bullet~400~{\rm Mah\bar{a}bh\bar{a}rata} \end{array}$  & 
 $\begin{array}{l} \bullet~312~{\rm Seleucidean~Era} \\ {\rm ~~~begins~at~}
 {\rm Babylon} \end{array}$  \\  \cline{2-4}
 & 5-th &   &   \\  \cline{2-4}
B.C. & 6-th & $\begin{array}{l} \bullet~560~{\rm Death~of} \\
 {\rm ~~~Gautama~Buddha} \end{array}$  & 
 $\begin{array}{l} \bullet~550~{\bf Pythagorus} \\ \bullet~560~{\rm Cyrus~}
 {\rm establishes} \\ {\rm ~~~Achaemenian~Empire} \\ {\rm ~~~in~Iran,} \\
 {\rm ~~~End~of~Assyrian} \\ {\rm ~~~Empire~at~Babylon,} \\
 {\rm ~~~End~of~Egyptian~Empire} \end{array}$  \\  \cline{2-4}
 & 7-th & $\begin{array}{l} \bullet~640~{\rm Birth~of} \\
 {\rm ~~~Gautama~Buddha} \end{array}$  &
 $\begin{array}{l} \bullet~650~{\rm Ashurbanipal~rules} \\
 {\rm ~~~at~Babylon} \end{array}$  \\  \cline{2-4}
 & 8-th &   & $\begin{array}{l} \bullet~728~{\rm Beginning~of} \\
 {\rm ~~~Assyrian~Empire~at} \\ {\rm ~~~Babylon,} \\ {\rm ~~~End~of} \\
 {\rm ~~~Chaldean~Empire} \\ {\rm ~~~at~Babylon,} \end{array}$ \\  \cline{2-4}
 & 9-th &   &   \\  \cline{2-4}
 & 10-th &   &   \\  \cline{2-4}
 & 11-th &   & $\begin{array}{l} \bullet~1100~{\rm ~~~End~of} \\
 {\rm ~~~New~Kingdom~at} \\ {\rm ~~~Egypt} \end{array}$  \\  \cline{2-4}
 & 12-th & $\begin{array}{l} \bullet~1200~{\rm Beginning~of} \\
 {\rm ~~~Ved\bar{a}\dot{n}ga~Jyoti\d{s}a} \end{array}$  &    \\  \cline{2-4}
 & $\cdots$ & $\cdots$ & $\cdots$ \\  \hline \hline
\end{tabular}
\pagebreak
\newpage 
Table \ref{histtab} (Continued) \\
\begin{tabular}{||l|l||l|l||}  \hline  \hline
{\sl Era\/} & {\sl Century\/} & {\sl Indian\/} & 
 {\sl West Asian\/}  \\  \hline  \hline
 & $\cdots$ & $\cdots$ & $\cdots$ \\  \cline{2-4} 
 & 13-th &   &    \\  \cline{2-4}
 & 14-th &   &    \\  \cline{2-4}
 & 15-th & $\begin{array}{l} \bullet~1500~{\rm \d{R}k-Veda} \\
 {\rm ~~~(terminus~ad~quem)} \end{array}$ & 
 $\begin{array}{l} \bullet~1500~{\rm End~of} \\ {\rm ~~~Middle~Kingdom} \\
 {\rm ~~~at~Egypt} \end{array}$ \\  \cline{2-4}
B.C. & 16-th &   &    \\  \cline{2-4}
 & 17-th &   &   \\  \cline{2-4}
 & 18-th &   &    \\  \cline{2-4}
 & 19-th &   & $\begin{array}{l} \bullet~1900~{\rm Hammurabi} \\
 {\rm ~~~establishes} \\ {\rm ~~~Chaldean~Empire} \\ {\rm ~~~at~Babylon}
 \end{array}$  \\  \cline{2-4}
 & 20-th & $\begin{array}{l} {\rm Harappan~civilization} \end{array}$ & 
   \\  \hline   \hline
\end{tabular}

\section{The {\sl Vedas} \label{vedas}}

\indent {\hskip 0.4cm} The earliest refernce to astronomical facts
are found in the Vedas, the scripture of the Hindus. There are 4 Vedas,
\d{R}k, S\={a}ma, Yaju\d{h} and Atharva. Each of these Vedas are
divided in 4 parts, Chanda (or Sa\.{m}hit\={a}), Br\={a}hma\d{n}a,
\={A}ra\d{n}yaka (or Upani\d{s}ad) and Ved\={a}\.{n}ga. Ved\={a}\.{n}ga
deals with Phonetics, Ritualistic literature, Grammar, Etymology,
Metrics and Astronomy. The Sa\.{m}hit\={a} part of \d{R}k-veda has
very scanty astronomical references. That of Yaju\d{h}  has slightly
more references. The Ved\={a}\.{n}gas (particularly of the K\d{r}\d{s}\d{n}a
or the Black Yaju\d{h}-veda), on the other hand, has a full-fledged
astronomy. The salient features of the Vedas are given in Table
\ref{vedastab}. This shows the range of academic activities pursued
by the ancient Indians.
\pagebreak
\newpage
\textheight 23.0cm
\begin{tab} 
{\rm The Vedas, their subdivisions and salient features.}
  \label{vedastab} 
\end{tab}
\begin{tabular}{cccccc} 
 &   &  $\begin{array}{c} The~Vedas\\ \downarrow \end{array}$  &   &   &
 \\  \cline{1-4}
 $\begin{array}{c} \downarrow \\ {\it \d{R}k\/} \\ {\rm divided~in} \\
 {\it 10~Ma\d{n}dalas\/} \end{array}$ &
 $\begin{array}{c} \downarrow \\ {\it S\bar{a}ma\/} \\ \\  \\  \end{array}$ &
 $\begin{array}{c} \downarrow \\ {\it Yaju\d{h}\/} \\  \\ \downarrow
 \end{array}$ &
 $\begin{array}{c} \downarrow \\ Atharva \\  \\  \\  \end{array}$ &   &
 \\  \cline{3-4}
  &   & $\begin{array}{c} \downarrow \\ {\it K\d{r}\d{s}\d{n}a\/} \\ (Black) \\
 \downarrow \end{array}$ &
  $\begin{array}{c} \downarrow \\ {\it \acute{S}ukla\/} \\ (White) \\   \\
 \end{array}$ &   &   \\  \cline{1-4}
 $\begin{array}{c} \downarrow \\ Chandas~{\rm or} \\ {\it Sa\dot{m}hit\bar{a}\/}
 \\ {\rm Collection} \\ {\rm of~hymns,} \\ {\rm prayers,} \\ {\rm incantations,}
 \\ {\rm benedictions,} \\ {\rm sacrificial} \\ {\rm formulas,} \\
 {\rm litanies} \end{array}$ &
 $\begin{array}{c} \downarrow \\ {\it Br\bar{a}hma\d{n}a\/} \\
 {\rm prose~texts} \\ {\rm on} \\ {\rm theological} \\ {\rm matters,} \\
 {\rm observations} \\ {\rm on~sacrifices} \\ {\rm and~their} \\
 {\rm mystical} \\ {\rm significances}  \end{array}$ &
 $\begin{array}{c} \downarrow \\ {\it \bar{A}ra\d{n}yaka\/}~{\rm or} \\
 {\it Upani\d{s}ad\/} \\ {\rm meditations} \\ {\rm of~forest} \\ 
 {\rm hermits} \\ {\rm and} \\ {\rm ascetics} \\ {\rm on~God,} \\
 {\rm the~world~\&} \\ {\rm mankind} \end{array}$ &
 $\begin{array}{c} \downarrow \\ {\it Ved\bar{a}\dot{n}ga\/} \\ {\rm or} \\
 {\it S\bar{u}tra\/} \\ \\ \\ \\ \\  \\ \\ \downarrow \end{array}$  &   & 
 \\ \hline
 $\begin{array}{c} \downarrow \\ {\it \acute{S}ik\d{s}\bar{a}\/} \\
 ({\rm Phonetics}) \\   \\  \\  \\  \\   \\  \\  \\  \\   \\ \\  \\  \\
 \end{array}$ &
 $\begin{array}{c} \downarrow \\ Kalpa \\ \left({\rm Ritualistic}\right. \\ 
 \left. {\rm literature}\right) \\ \\ \\ \\ \\ \\ \\ \\ \\ \\ \\  \\
 \end{array}$ &
 $\begin{array}{c} \downarrow \\ {\it Vy\bar{a}kara\d{n}a\/} \\
 ({\rm Grammar}) \\ {\rm e.g.~} \\ P\bar{a}\d{n}ini{\rm 's} \\
 {\it A\d{s}\d{t}\bar{a}dhy\bar{a}y\bar{i}\/} \\ \\ \\ \\ \\ \\ \\ \\  \\  \\
 \end{array}$ &
 $\begin{array}{c} \downarrow \\ Nirukta \\ ({\rm Etymology}) \\ 
 {\rm e.g.~of~} \\ {\it Y\bar{a}\d{s}ka\/} \\ \\ \\ \\ \\ \\ \\  \\  \\  \\  \\
 \end{array}$ &
 $\begin{array}{c} \downarrow \\ Chandas \\ ({\rm Metrics}) \\
 {\rm e.g.~of~} \\ {\it Pi\dot{n}gala\/}   \\ \\ \\ \\ \\ \\ \\ \\ \\  \\  \\
 \end{array}$ &
 $\begin{array}{c} \downarrow \\ {\it Jyoti\d{s}a\/} \\({\rm Astro-}\\
 {\rm -nomy}) \\ {\rm ~e.g.~of~} \\ Lagadha \\ {\it \d{R}k-\/} \\
 {\it Jyoti\d{s}a\/} \\ {\rm has~}36 \\ {\rm verses} \\ {\rm and} \\
 {\it Yaju\d{h}-\/} \\ {\it Jyoti\d{s}a\/} \\ {\rm has~}43 \\ {\rm verses}
 \end{array}$
\end{tabular}

\section{The {\sl Siddh\={a}nta}s  \label{sid}}

\indent {\hskip 0.4cm} Hindu astronomy gradually developed from
the Ved\={a}\.{n}ga jyoti\d{s}a to what are known as the Siddh\={a}ntas.
There are 5 Siddh\={a}ntas: Pait\={a}maha, V\={a}\'{s}i\d{s}\d{t}ha,
Romaka, Pauli\'{s}a and S\={u}rya. The last one is the latest and the
most developed. Texts of all the siddh\={a}ntas are not available,
but Var\={a}hamihira in the 6th century A.D. has described them in his
Pa\~{n}ca-Siddh\={a}ntik\={a}. Romaka and Pauli\'{s}a siddh\={a}ntas show
influence of Babylonian and Egyptian astronomy. One of the earliest
proponent of the S\={u}rya Siddh\={a}nta was \={A}ryabha\d{t}a. But
with time it was upgraded with accumulation of new knowledge. Solar,
lunar and stellar motions were accurately described in this Siddh\={a}nta.
A study of this Siddh\={a}nta shows that Hindu astronomers knew about 
precession of the Equinoxes, but they were lamentably unaware of the 
difference between the Sidereal and the Tropical years. This is not so 
amazing, since even European astronomers became aware of this difference 
only in the 17th century A.D. after Galileo, Kepler and Newton. The 5 
Siddh\={a}ntas are summarized in Table \ref{sidtab}.
\pagebreak
\newpage
\begin{tab} 
{\rm Different Siddh\={a}nta Jyoti\d{s}as and their salient features.}
   \label{sidtab}
\end{tab}
\begin{tabular}{ccccc}
  &   & $\begin{array}{c} The~Siddh\bar{a}ntas \\  \downarrow \end{array}$ &
   &   \\  \hline
 $\begin{array}{c} \downarrow \\ {\it Pait\bar{a}maha\/} \\ {\rm ascribed~to}
 \\ {\rm Brahm\bar{a}~and} \\ {\rm described~in} \\ {\rm 5~stanzas~in} \\
 {\it Pa\tilde{n}ca-\/} \\ {\it Siddh\bar{a}ntik\bar{a}\/} \\ {\rm of} \\
 {\rm Var\bar{a}hamihira} \\  \\  \\  \\  \\  \\  \\  \\  \\  \\  \\ 
 \\  \\ \\  \\  \\  \\  \\  \\  \\  \\  \\  \\  \\  \\ \\  \\  \\  \\  \\
 \end{array}$ &
 $\begin{array}{c} \downarrow \\ {\it V\bar{a}\acute{s}i\d{s}\d{t}ha\/} \\
 {\rm ascribed~to} \\ {\rm Va\acute{s}i\d{s}\d{t}ha~and} \\ {\rm revealed} \\
 {\rm by~him~to} \\ {\rm M\bar{a}\d{n}davya} \\ {\rm and} \\
 {\rm described~in} \\ 13~{\rm couplets} \\ {\rm by} \\
 {\rm Var\bar{a}hamihira;} \\ {\rm method~of} \\ {\rm calculating} \\
 Tithi~{\rm and} \\ {\it Nak\d{s}atra\/}, \\ {\rm mentions} \\
 {\it R\bar{a}\acute{s}i\/} \\ {\rm and} \\ Lagna \\  \\  \\  \\  \\  \\  \\ 
 \\ \\  \\ \\  \\  \\  \\  \\  \\  \\  \\  \\  \\   \end{array}$ &
 $\begin{array}{c} \downarrow \\ Romaka \\ {\rm revealed~by} \\
 {\rm Vi\d{s}\d{n}u} \\
 {\rm to~\d{R}\d{s}i} \\ {\rm Roma\acute{s}a;} \\ Yuga~=~ \\ 19\times 5\times
 30~yrs, \\ {\rm incorporates} \\ 19~yr~{\rm cycle~of} \\ {\rm Babylon~and} \\
 5~yr~Yuga~{\rm of} \\ {\it Ved\bar{a}\dot{n}ga\/} \\ {\it Jyoti\d{s}a\/} \\
 \\  \\ \\  \\ \\  \\  \\  \\  \\  \\  \\  \\ \\   \\  \\   \\  \\   \\ 
 \\  \\  \\  \\  \\  \\   \\  \end{array}$ &
 $\begin{array}{c} \downarrow \\ {\it Pauli\acute{s}a\/} \\ {\rm ascribed~to}
 \\ {\rm the~sage} \\ {\rm Pulastya;} \\ {\rm refers~to} \\ Yavanapura, \\
 {\rm Longitudes~of} \\ {\rm Ujjain\bar{i}~\&} \\ {\rm Benares} \\
 {\rm are~given} \\ {\rm with~respect} \\ {\rm to} \\ {\rm Alexandria,} \\
 {\rm Lords~of} \\ {\rm the~days} \\ {\rm are} \\ {\rm mentioned} \\
 {\rm like~the} \\ {\rm Iranian} \\ {\rm calendar,} \\ {\rm but~the} \\
 {\rm names~are} \\ {\rm Indian} \\   \\   \\  \\   \\  \\  \\   \\  \\  \\
 \\  \\  \\  \\  \\   \\   \end{array}$ &
 $\begin{array}{c} \downarrow \\ {\it S\bar{u}rya\/} \\ {\rm Revealed} \\
 {\rm by~the} \\ {\rm Sun~god} \\ {\rm to~Asura} \\ {\rm Maya,} \\ {\rm who} \\
 {\rm propounded} \\ {\rm it~to} \\ {\rm the~\d{R}\d{s}is.} \\ {\rm It~has} \\
 {\rm 500~verses~in} \\ {\rm 14~Chapters,} \\ {\rm does~not} \\
 {\rm show~any} \\ {\rm influence~of} \\ {\rm Ptolemy's} \\ {\rm Almagest,} \\
 {\rm does~not} \\ {\rm mention} \\ {\rm Alexandria,} \\ {\rm Longitude} \\
 {\rm of~Ujjain\bar{i}} \\ {\rm is~the} \\ {\rm standard} \\ {\rm Meridian,} \\
 {\rm it~is} \\ {\rm continually} \\ {\rm updated} \\ {\rm like~the} \\
 {\rm Landolt-} \\ {\rm Bernstein} \\ {\rm tables,} \\ {\rm earliest} \\
 {\rm version} \\ {\rm is~by} \\ {\rm \bar{A}ryabha\d{t}a} \\
 476~-~523 {\rm A.D.}    \end{array}$  \\ 
\end{tabular}

\section{Comparison of different Astronomical systems of India before 1200 A.D.
   \label{astron}}
\textheight 21.0cm

\indent {\hskip 0.4cm} In Table \ref{astrotab} we have summarized
astronomical knowledge of the Hindu astronomers found in the 4 systems:
\d{R}k-Sa\.{m}hit\={a}, Yaju\d{h}-Sa\.{m}hit\={a}, Ved\={a}\.{n}ga
Jyoti\d{s}a and S\={u}rya Siddh\={a}nta. This knowledge primarily
encompasses the following subjects: Lunar asterism, Lunar day (Tithi),
Month \& Year (both Lunar and Solar), Seasons \& Year (\d{R}tu), and
longer period of time (Yuga).

In the \d{R}k-Sa\.{m}hit\={a}, scanty though the knowledge is, we
find mentions of lunar asterism, some specific fixed stars, lunar
zodiac, lunar year of 360 days and solar year of 366 days, 3 Seasons
each of 4 months (C\={a}turm\={a}sya) and a 5 year Yuga. It is
interesting to note that in the list of Seasons, the Rainy Season
was not mentioned. This is possibly because the Hindu astronomers
did not come across this season at the place where they were studying
astronomy. This indicates that they were located only in the extreme
North-Western corner of the country (Tak\d{s}a\'{s}il\={a}) and
Afghanistan (Balkh).

During the time of the Yaju\d{h}-Sa\.{m}hit\={a} astronomy was more
developed. Full list of the Lunar asterism is given though it
started with K\d{r}ttik\={a} instead of A\'{s}vi\d{n}\={i} as in
later time. Lunar fortnight (Pak\d{s}a) of 15 lunar days
(Pa\~{n}cada\'{s}\={i} tithi) was mentioned and gradually the
concept of lunar month came into vogue. Northern and Southern
motions of the Sun and the 6 Seasons connected with the motion of
the sun along the Ecliptic was established knowledge. The Spring
(Vasanta \d{R}tu) consisting of 2 months was symmetrically placed
about the Vernal Equinox (V.E.) point and the Solar year started 
when the Sun was $30^{\circ}$ before the V.E.

Ved\={a}\.{n}ga Jyoti\d{s}a was a completely developed system with 
the numerical values of astronomical data comparing favorably with
modern values. Evidence of the knowledge of the Sidereal Day, the
Solstices, ratio of durations of day \& night on the Summer Solstice,
Lunar months, Tropical Solar Year, and longer periods of time is
found in Ved\={a}\.{n}ga Jyoti\d{s}a. During the time of the
Ved\={a}\.{n}ga Jyoti\d{s}a, Winter Solstice was at the
Sravi\d{s}\d{t}h\={a} nak\d{s}atra (Delphini) at the longitude of
$270^{\circ}$. Sravi\d{s}\d{t}h\={a} is now at $316^{\circ}~~41^{\prime}$.
This fixes the time of the Ved\={a}\.{n}ga Jyoti\d{s}a around
1400 B.C. Similar calculation fixes the time of compilation of the
Mah\={a}bh\={a}rata as 450 B.C.

S\={u}rya Siddh\={a}nta has precise and logical definitions of the
Lunar asterism, Tithi (Lunar Day), Solar Month, Tropical (S\={a}vana)
year and the 6 Seasons (\d{R}tu). The effect of the precession of
the Equinoxes was also noted.
\pagebreak
\newpage
\begin{tab}     
{\rm Salient features of Hindu Astronomy in different ages.}
   \label{astrotab}
\end{tab}
\begin{tabular}{||l||l|l|l|l|||}  \hline  \hline
 $\begin{array}{c} Astro- \\ ~~nomical \\  features \end{array}$ & 
 $\begin{array}{c} \d{R}k \\ Sa\dot{m}hit\bar{a} \end{array}$   &
 $\begin{array}{c} Yaju\d{h} \\ Sa\dot{m}hit\bar{a} \end{array}$  &
 $\begin{array}{c} Ved\bar{a}\dot{n}ga \\ Jyoti\d{s}a \end{array}$ &
 $\begin{array}{c} S\bar{u}rya \\ Siddh\bar{a}nta \end{array}$ 
 \\  \hline  \hline
$\begin{array}{l} {\rm Lunar} \\ {\rm Asterism} \\ ({\rm Nak\d{s}atra})
 \end{array}$  &
$\begin{array}{l} \bullet~{\rm Custom~of} \\ {\rm ~~mentioning}
 \\ {\rm ~~Nak\d{s}atra} \\ {\rm ~~perhaps} \\ {\rm ~~started} \\ 
 {\rm ~~~here} \\ \bullet~{\rm Days~were} \\ {\rm ~~designated} \\ 
 {\rm ~~by~lunar} \\ {\rm ~~asterism} \\  \bullet~{\rm Agh\bar{a}~\&} \\ 
 {\rm ~~Arjun\bar{i}} \\ {\rm ~~were} \\ {\rm ~~mentioned} \end{array}$ & 
$\begin{array}{l} \bullet~{\rm Full~list} \\ {\rm ~~of~lunar} \\
 {\rm ~~asterism~is} \\ {\rm ~~given} \\ \bullet~{\rm Lunar} \\
 {\rm ~~asterism} \\ {\rm ~~starts~with} \\ {\rm ~~K\d{r}ttik\bar{a}} 
 \end{array}$  &
 $\begin{array}{l} \bullet~{\rm Lunar~asterism} \\
 {\rm ~~starts~at~K\d{r}ttik\bar{a},} \\ {\rm ~~most~probably~}\Upsilon \\
 \bullet~{\rm Days~were~named} \\ {\rm ~~after~Nak\d{s}atra} \\
 \bullet~1~{\rm N\bar{a}k\d{s}atra~day} \\ {\rm ~~was}~1.011608 \\
 {\rm ~~Solar~day,} \\ {\rm ~~in~reality~it~is} \\ ~~1.011913 \\
 {\rm ~~Solar~day} \\ \bullet~{\rm W.S.~was~at} \\
 {\rm ~~\acute{S}ravi\d{s}\d{t}h\bar{a}~at} \\ {\rm ~~longitude~}270^{\circ};
 \\ {\rm ~~it~is~now~at} \\ ~~316^{\circ}~41^{\prime}; \\
 {\rm ~~solstice~takes} \\ ~~72~{\rm yrs~to} \\ {\rm ~~retrograde} \\
 {\rm ~~through~}1^{\circ}; \\ {\rm ~~if~}\alpha~-~{\rm Delphini,} \\
 {\rm ~~then~time~is} \\ ~~1413~{\rm B.C.,} \\
 {\rm ~~if~}\beta~-~{\rm Delphini,} \\ {\rm ~~then~time~is} \\
 ~~1338~{\rm B.C.} \\ \bullet~{\rm In~Mah\bar{a}bh\bar{a}rata} \\
 {\rm ~~W.S.~is~at~\acute{S}rava\d{n}\bar{a},} \\ {\rm ~~time~is~450~B.C.}
 \end{array}$ & 
$\begin{array}{l} \bullet~{\rm Lunar} \\ {\rm ~~asterism} \\
 {\rm ~~starts~at} \\ {\rm ~~A\acute{s}vin\bar{i}} \\ ~~(\alpha-~{\rm or~}
 \beta- \\ {\rm ~~~~Arietis}), \\ ~~\Upsilon~{\rm was~~at} \\ {\rm ~~Revati~} \\
 ~~(\zeta-{\rm Piscium});\\ {\rm ~~now}~\Upsilon{\rm ~has} \\
 {\rm ~~shifted~by} \\ ~~\sim~19^{\circ}~{\rm from} \\ ~~\zeta-~{\rm Piscium} \\
 \bullet~{\rm At~Var\bar{a}ha-} \\ {\rm ~~mihira's} \\ {\rm ~~time~S.S.} \\
 {\rm ~~has~moved} \\ ~~\frac{1}{2}~{\rm of~A\acute{s}le\d{s}\bar{a}}~+~ \\
 {\rm ~~ Pu\d{s}y\bar{a},~time} \\ {\rm ~~is~1500~yrs} \\
 {\rm ~~after~Ved\bar{a}\dot{n}ga} \\ {\rm ~~Jyoti\d{s}a} \\ 
 \bullet~{\rm No~knowledge} \\ {\rm ~~of~precession} \\ {\rm ~~of~the} \\
 {\rm ~~Equinoxes,} \\ {\rm ~~but~its~effect} \\ {\rm ~~has~been} \\
 {\rm ~~observed} \end{array}$ \\  \hline 
 $\cdots$ & $\cdots$ & $\cdots$ & $\cdots$ & $\cdots$ \\  \hline
\end{tabular}
\pagebreak
\newpage
Table \ref{astrotab} (Continued) \\
\begin{tabular}{||l||l|l|l|l|||}  \hline  \hline
 $\begin{array}{c} Astro- \\ ~~nomical \\  features \end{array}$ & 
 $\begin{array}{c} \d{R}k \\ Sa\dot{m}hit\bar{a} \end{array}$   &
 $\begin{array}{c} Yaju\d{h} \\ Sa\dot{m}hit\bar{a} \end{array}$  &
 $\begin{array}{c} Ved\bar{a}\dot{n}ga \\ Jyoti\d{s}a \end{array}$ &
 $\begin{array}{c} S\bar{u}rya \\ Siddh\bar{a}nta \end{array}$ 
 \\  \hline  \hline
 $\cdots$ & $\cdots$ & $\cdots$ & $\cdots$ & $\cdots$ \\  \hline
$\begin{array}{l} {\rm Lunar} \\ {\rm day} \\ ({\rm Tithi}) \end{array}$ &
$\begin{array}{l} \bullet~{\rm No} \\ {\rm ~~reference} \\
 \bullet~{\rm Perhaps} \\ {\rm ~~individual} \\ {\rm ~~days~were} \\
 {\rm ~~denoted} \\ {\rm ~~by~the} \\ {\rm ~~Nak\d{s}atra,} \\
 \bullet~{\rm Lunar} \\ {\rm ~~zodiac} \\ {\rm ~~was~known} \end{array}$ &
$\begin{array}{l} \bullet~{\rm Pa\tilde{n}cada\acute{s}\bar{i}} \\
 {\rm ~~tithi} \\ \bullet~{\rm Knowledge} \\ {\rm ~~of~Pak\d{s}a} \\
 \bullet~{\rm Tithi} \\ {\rm ~~was~from} \\ {\rm ~~moonrise} \\
 {\rm ~~to~moonrise} \\ {\rm ~~during} \\ {\rm ~~K\d{r}\d{s}\d{n}a~pak\d{s}a} \\
 {\rm ~~and~from} \\ {\rm ~~moonset~to} \\ {\rm ~~moonset} \\ {\rm ~~during} \\
 {\rm ~~\acute{S}ukla~pak\d{s}a} \end{array}$ & 
$\begin{array}{l} \bullet~{\rm Subdivision} \\ {\rm ~~of~day~was} \\
 {\rm ~~measured~by} \\ {\rm ~~Clepsydra} \\ \bullet~{\rm Ratio~of} \\
 {\rm ~~day~\&~night} \\ {\rm ~~on~S.S.~is}~\frac{3}{2}, \\
 {\rm ~~this~is} \\ {\rm ~~true~for} \\ {\rm ~~latitude} \\ 
 ~~35^{\circ}~{\rm N} \\ \bullet~{\rm Tithi~is~}\frac{1}{30} \\
 {\rm ~~of~lunar} \\ {\rm ~~month~=} \\ ~~0.983871 \\ {\rm ~~days,} \\
 {\rm ~~correct} \\ {\rm ~~value~is} \\ ~~0.984353 \\ {\rm ~~days} \\
 \bullet~{\rm Day~is} \\ {\rm ~~reckoned} \\ {\rm ~~by~tithi} \end{array}$  &
$\begin{array}{l} \bullet~{\rm Tithi~is} \\ {\rm ~~complete~when} \\
 {\rm ~~the~moon~is} \\ {\rm ~~ahead~of} \\ {\rm ~~the~sun~by} \\
 ~~12^{\circ}{\rm ~or~its} \\ {\rm ~~multiple,~tithi} \\ {\rm ~~varies~from} \\
 ~~26~{\rm hr}~47~{\rm min~to} \\ ~~19~{\rm hr}~59~{\rm min} \\
 \bullet~{\rm If~a~tithi} \\ {\rm ~~extends~to} \\ ~~2~{\rm solar~days,} \\
 {\rm ~~then~both} \\ {\rm ~~the~days} \\ {\rm ~~are~assigned} \\
 {\rm ~~the~same} \\ {\rm ~~tithi} \end{array}$  \\  \hline 
 $\cdots$ & $\cdots$ & $\cdots$ & $\cdots$ & $\cdots$ \\  \hline
\end{tabular}
\pagebreak
\newpage
Table \ref{astrotab} (continued) \\
\begin{tabular}{||l||l|l|l|l|||}  \hline  \hline
 $\begin{array}{c} Astro- \\ ~~nomical \\  features \end{array}$ & 
 $\begin{array}{c} \d{R}k \\ Sa\dot{m}hit\bar{a} \end{array}$   &
 $\begin{array}{c} Yaju\d{h} \\ Sa\dot{m}hit\bar{a} \end{array}$  &
 $\begin{array}{c} Ved\bar{a}\dot{n}ga \\ Jyoti\d{s}a \end{array}$ &
 $\begin{array}{c} S\bar{u}rya \\ Siddh\bar{a}nta \end{array}$ 
 \\  \hline  \hline
 $\cdots$ & $\cdots$ & $\cdots$ & $\cdots$ & $\cdots$ \\  \hline
$\begin{array}{l} {\rm Month} \\ {\rm \&~Year} \end{array}$ &
$\begin{array}{l} \bullet~{\rm Year~has} \\ {\rm ~~12~months} \\
 {\rm ~~each~of} \\ {\rm ~~30~days} \\ ~~\left(12~{\rm spoke-}\right. \\
 {\rm ~~boards,~\&} \\ ~~\left. 360~{\rm \acute{s}a\dot{n}kus}\right) \\
 \bullet~{\rm Mentions} \\ {\rm ~~Citr\bar{a}} \\ ~~(\alpha-{\rm Virginis)} \\
 {\rm ~~\&~Agh\bar{a}} \\ ~~(\alpha-{\rm Leonis}) \end {array}$ & 
$\begin{array}{l} \bullet~{\rm Mentions} \\ {\rm ~~lunar} \\ {\rm ~~mansions}\\
 {\rm ~~with~their} \\ {\rm ~~presiding} \\ {\rm ~~deities} \\
 \bullet~{\rm Lunar} \\ {\rm ~~months} \\ {\rm ~~came} \\ {\rm ~~gradually,} \\
 {\rm ~~generally} \\ {\rm ~~pak\d{s}a~is} \\ {\rm ~~mentioned} \end{array}$  &
$\begin{array}{l} \bullet~{\rm Does~not} \\ {\rm ~~mention} \\
 {\rm ~~zodiac} \\ {\rm ~~signs} \\ \bullet~{\rm P\bar{u}r\d{n}im\bar{a}nta}\\
 {\rm ~~lunar} \\ {\rm ~~months} \\ {\rm ~~are~named} \\ {\rm ~~after~the} \\
 {\rm ~~Nak\d{s}atras} \end{array}$   & 
$\begin{array}{l} \bullet~{\rm Defines} \\ {\rm ~~only~astro-} \\
 {\rm ~~nomical} \\ {\rm ~~solar} \\ {\rm ~~month} \\
 \bullet~{\rm 1~Solar} \\ {\rm ~~month~=} \\ {\rm ~~time~of} \\
 {\rm ~~passage} \\ {\rm ~~of~the} \\ {\rm ~~Sun} \\ {\rm ~~through} \\
 ~~30^{\circ}~{\rm of~its} \\ {\rm ~~orbit~=} \\ ~~30.43823 \\ {\rm ~~days;} \\
 {\rm ~~modern} \\ {\rm ~~value~=} \\ ~~30.43685 \\ {\rm ~~days} \\
 \bullet~{\rm 1~Solar} \\ {\rm ~~year~=} \\ {\rm ~~12~Solar} \\
 {\rm ~~months} \end{array}$
 \\  \hline
 $\cdots$ & $\cdots$ & $\cdots$ & $\cdots$ & $\cdots$ \\  \hline
\end{tabular}
\pagebreak
\newpage
\textheight 23.0cm
Table \ref{astrotab} (Continued) \\
\begin{tabular}{||l||l|l|l|l|||}  \hline  \hline
 $\begin{array}{c} Astro- \\ ~~nomical \\  features \end{array}$ & 
 $\begin{array}{c} \d{R}k \\ Sa\dot{m}hit\bar{a} \end{array}$   &
 $\begin{array}{c} Yaju\d{h} \\ Sa\dot{m}hit\bar{a} \end{array}$  &
 $\begin{array}{c} Ved\bar{a}\dot{n}ga \\ Jyoti\d{s}a \end{array}$ &
 $\begin{array}{c} S\bar{u}rya \\ Siddh\bar{a}nta \end{array}$ 
 \\  \hline  \hline
 $\cdots$ & $\cdots$ & $\cdots$ & $\cdots$ & $\cdots$ \\  \hline
$\begin{array}{l} {\rm Seasons} \\ {\rm \&~Year} \end{array}$ &
$\begin{array}{l} \bullet~{\rm 3~seasons} \\ {\rm ~~(navels),} \\
 {\rm ~~C\bar{a}turm\bar{a}sya,} \\ {\rm ~~each~of} \\ {\rm ~~4~months}, \\
 {\rm ~~Gr\bar{i}\d{s}ma} \\ {\rm ~~(Summer),} \\ {\rm ~~\acute{S}arad} \\
 {\rm ~~(Autumn),} \\ {\rm ~~Hemanta} \\ {\rm ~~(Winter);} \\ 
 {\rm ~~Var\d{s}\bar{a}} \\ {\rm ~~(Rainy)} \\ {\rm ~~is~not} \\
 {\rm ~~mentioned} \\ \bullet~{\rm Lunar} \\ {\rm ~~year~of} \\
 {\rm ~~360~days,} \\ {\rm ~~12~lunar} \\ {\rm ~~months~and} \\
 {\rm ~~1~inter-} \\ {\rm ~~calary} \\ {\rm ~~month~of} \\
 {\rm ~~12~tithis} \\ \bullet~{\rm Solar} \\ {\rm ~~year~of} \\
 {\rm ~~366~days} \end{array}$ &  
$\begin{array}{l} \bullet~{\rm Mentions} \\ {\rm ~~Uttar\bar{a}ya\d{n}a} \\
 {\rm ~~\&~Dak\d{s}i\d{n}\bar{a}yana} \\ {\rm ~~of~the~Sun} \\
 \bullet~{\rm The~Solar~year} \\ {\rm ~~has~6~Seasons:} \\
 {\rm ~~i)~Vasanta} \\ {\rm ~~~(Spring)~=} \\ {\rm ~~~Madhu~\&} \\
 {\rm ~~~M\bar{a}dhava} \\ {\rm ~~ii)~Gr\bar{i}\d{s}ma} \\
 {\rm ~~(Summer)~=} \\ {\rm ~~~\acute{S}ukra~\&~\acute{S}uci} \\
 {\rm ~~iii)~Var\d{s}\bar{a}~(Rains)} \\ {\rm ~~~=Nabha~\&} \\
 {\rm ~~~Nabhasya} \\ {\rm ~~iv)~\acute{S}arad~(Autumn)} \\
 {\rm ~~~=~I\d{s}a~\&~\bar{U}rja} \\ {\rm ~~v)~Hemanta} \\
 {\rm ~~~(Early~Winter)~=} \\ {\rm ~~~Sahas~\&} \\ {\rm ~~~Sahasya} \\
 {\rm ~~vi)~\acute{S}\bar{i}ta~(Winter)~=} \\ {\rm ~~~Tapas~\&~Tapasya;} \\
 {\rm ~~each~Season} \\ {\rm ~~covers}~60^{\circ}~{\rm of} \\
 {\rm ~~the~Sun's~motion} \\ {\rm ~~along~the} \\ {\rm ~~Ecliptic;} \\
 {\rm ~~solar~months} \\ {\rm ~~are~seldom} \\ {\rm ~~mentioned} \\
 \bullet~{\rm Tropical~Solar} \\ {\rm ~~year~\&~Vasanta} \\
 {\rm ~~starts~at~}-30^{\circ}~{\rm of} \\ {\rm ~~the~Ecliptic,} \\
 \end{array}$ & 
$\begin{array}{l} \bullet~{\rm In~Br\bar{a}hm\bar{i}} \\ {\rm ~~inscrip-} \\
 {\rm ~~tions} \\ {\rm ~~of~the} \\ {\rm ~~Ku\d{s}\bar{a}\d{n}as} \\
 {\rm ~~month} \\ {\rm ~~names} \\ {\rm ~~are} \\ {\rm ~~mostly} \\
 {\rm ~~seasonal~:} \\ {\rm ~~Gr\bar{i}\d{s}ma,} \\ {\rm ~~Var\d{s}\bar{a},} \\
 {\rm ~~Hemanta} \\ \bullet~{\rm The~pak\d{s}a} \\ {\rm ~~is~not} \\
 {\rm ~~mentioned} \\ \bullet~{\rm Astro-} \\ {\rm ~~nomical} \\
 {\rm ~~year} \\ {\rm ~~starts} \\ {\rm ~~when} \\ {\rm ~~the~Sun} \\
 {\rm ~~crosses} \\ {\rm ~~the~V.E.,} \\ {\rm ~~the~civil} \\
 {\rm ~~year} \\ {\rm ~~starts} \\ {\rm ~~on~the} \\ {\rm ~~next} \\
 {\rm ~~day} \end{array}$  & 
$\begin{array}{l} \bullet~6~{\rm Seasons} \\ {\rm ~~(\d{R}tu)~in} \\
 {\rm ~~Tropical} \\ {\rm ~~Year;} \\ {\rm ~~each} \\ {\rm ~~season} \\
 {\rm ~~consists} \\ {\rm ~~of~2~zodiac} \\ {\rm ~~signs;} \\
 {\rm ~~each} \\ {\rm ~~zodiac} \\ {\rm ~~sign} \\ {\rm ~~covers} \\
 ~~30^{\circ}~{\rm of~the} \\ {\rm ~~Ecliptic} \\ \bullet~{\rm Tropical} \\
 {\rm ~~Year~=} \\ ~~365.25875 \\ {\rm ~~days} \\ {\rm ~~(Var\bar{a}ha-} \\
 {\rm ~~~mihira),} \\ ~~365.8756 \\ {\rm ~~days} \\ {\rm ~~(Current} \\
 {\rm ~~~~~~S.S),} \\ ~~365.242196 \\ {\rm ~~days} \\ {\rm ~~(Modern} \\
 {\rm ~Astronomy)} \\ \bullet~{\rm Solar} \\ {\rm ~~Year} \\ {\rm ~~starts~at}
 \\ {\rm ~Vai\acute{s}\bar{a}kha,} \\ {\rm ~~Lunar} \\ {\rm ~~Year} \\
 {\rm ~~starts~at} \\ {\rm ~~Caitra} \end{array}$\\  \hline
\end{tabular}
\pagebreak
\newpage
\textheight 21.0cm
Table \ref{astrotab} (Continued) \\
\begin{tabular}{||l||l|l|l|l|||}  \hline  \hline
 $\begin{array}{c} Astro- \\ ~~nomical \\  features \end{array}$ & 
 $\begin{array}{c} \d{R}k \\ Sa\dot{m}hit\bar{a} \end{array}$   &
 $\begin{array}{c} Yaju\d{h} \\ Sa\dot{m}hit\bar{a} \end{array}$  &
 $\begin{array}{c} Ved\bar{a}\dot{n}ga \\ Jyoti\d{s}a \end{array}$ &
 $\begin{array}{c} S\bar{u}rya \\ Siddh\bar{a}nta \end{array}$ 
 \\  \hline  \hline
 $\cdots$ & $\cdots$ & $\cdots$ & $\cdots$ & $\cdots$ \\  \hline
$\begin{array}{l} {\rm Longer} \\ {\rm Periods} \\ {\rm of~time} \end{array}$ &
$\begin{array}{l} \bullet~{\rm Yuga} \\ ~~(\equiv {\rm ~Saros~of} \\
 {\rm ~~Chaldean} \\ {\rm ~Astronomy)} \\ \bullet~{\rm Perhaps} \\
 {\rm ~~a~5~years} \\ {\rm ~~Yuga} \end{array}$ &   & 
$\begin{array}{l} \bullet{\rm Yuga} \\ ~~=~1830 \\ {\rm ~~s\bar{a}vana~days} \\
 ~~=~1860 \\ {\rm ~~~~~tithis} \\ ~~=~62~{\rm lunar} \\ {\rm ~~~~~months} \\
 ~~=~60~{\rm solar} \\ {\rm ~~~~~months} \\ ~~=~{\rm 67~n\bar{a}k\d{s}atra} \\
 {\rm ~~~~~months} \\ \bullet~{\rm Yuga} \\ {\rm ~~begins} \\ {\rm ~~at~W.S.} \\
 {\rm ~~with~the} \\ {\rm ~~Sun~and} \\ {\rm ~~the~moon} \\ 
 {\rm ~~at~Dhani\d{s}\d{t}h\bar{a}} \\ \bullet~{\rm Mah\bar{a}yuga} \\
 ~~4.32\times~10^6 \\ {\rm ~~years} \end{array}$  &
$\begin{array}{l} \bullet~{\rm Kalpa} \\ ~~4.32\times~10^9 \\ {\rm ~~years}
 \end{array}$  \\ \hline \hline
\end{tabular}

\section{Comments on Astrology \label{astrol}}

\indent {\hskip 1.0cm} Unlike in Babylonian civilization, astrology
had no place in Hindu astronomy. This is perhaps because of Gautama
Buddha's strictures against astrological predictions of doom and
disorder. This is found in the Buddhist scripture
 D\={i}gha Nik\={a}ya, Vol 1, p. 65, (Pali Text Book Society) 

Some br\={a}hma\d{n}as and \'{s}rama\d{n}as earn their livelihood by
taking to beastly professions and eating food brought to them out of
fear. They say: "There will be a solar eclipse, a lunar eclipse,
occultation of the stars, the sun and the moon will move in the
correct direction, in the incorrect direction, the nak\d{s}atras
will move in the correct path, in the incorrect path, there will be
precipitation of meteors, burning of the cardinal directions (?),
earthquakes, roar of heavenly war-drums, the sun, the moon and
the stars will rise and set wrongly producing wide distress amongst
all beings, etc. "

 Kautily\={i}ya Artha\'{s}\={a}stra also has the following stricture
against astrological predictions.

"The objective ($\equiv$ artha) eludes the foolish man ($\equiv$ b\={a}lam)
who enquires too much from the stars. The objective should be the
nak\d{s}atra of the objective, of what avail are the stars."

The practice of astrological prediction of human fate was perhaps 
imported from Babylonian civilization by the \'{S}akas in the 1st
century B.C. It is also to be noted that in the Hindu books of
knowledge Astronomy was called Jyoti\d{s}a whereas Astrology was
called S\={a}mudrika Jyoti\d{s}a. The epithet S\={a}mudrika indicates
that this branch of knowledge came from beyond the boundaries of the
country.


\begin{thebibliography}{99}
\bibitem{albiruni} al-B\={i}r\={u}n\={i}, {\sl Kit\={a}b-ul-Hind\/} and
 {\sl \={A}th\={a}r-al-B\={a}quia\/}, English translation 
 {\sl Alberuni's India\/} by E. C. Sachau, (London, 1914)
\bibitem{aryabhat} \={A}ryabha\d{t}a, {\sl \={A}ryabha\d{t}iya\/},
 English translation by P. C. Sengupta, (University of Calcutta,
 Calcutta)
\bibitem{bartholomew} John Bartholomew, {\sl The Oxford School Atlas\/}, 
  17-th Edition, (Oxford University Press, Oxford, 1952)
\bibitem{bhaskara1} Bh\={a}skar\={a}c\={a}rya, {\sl Siddh\={a}nta
 \'{S}iroma\d{n}i\/}, Edited by B. D. Sastri, (Benares, 1929)
\bibitem{bhaskara2} Bh\={a}skar\={a}c\={a}rya, {\sl Gol\={a}dhy\={a}ya\/},
 Edited by G. D. Sastri, (Benares, 1929)
\bibitem{bhattapal} Bha\d{t}\d{t}ap\={a}la, {\sl Commentary on
 B\d{r}hajj\={a}taka of Var\={a}hamihira\/}, Edited by Sita Ram Jha,
 (Benares, 1934)
\bibitem{diksit} S. B. Diksit, {\sl Bharatiya Jyoti\d{h}\'{s}\={a}stra\/}
\bibitem{surya} P. L. Gangoly (Editor), {\sl S\={u}rya Siddh\={a}nta\/},
 Reprint of the English translation with Notes and Appendix by E. Burgers,
 (University of Calcutta, Calcutta, 1936)
\bibitem{readers} {\sl Great World Atlas\/}, 3-rd Edition, (Reader's Digest
  Association, London, 1977)
\bibitem{gutzwiller} Martin C. Gutzwiller, Rev. Mod. Phys. {\bf 70}, 589(1998)
\bibitem{majumdar3} R. C. Majumdar (General Editor), {\sl The History and
  Culture of the Indian People\/}, Vol. III, ({\sl The Classical Age\/}),
  4-th Edition, (Bharatiya Vidya Bhavan, Bombay, 1988)
\bibitem{majumdar4} R. C. Majumdar (General Editor), {\sl The History and
  Culture of the Indian People\/}, Vol. IV, ({\sl The Age of Imperial
  Kanauj\/}), 4-th Edition, (Bharatiya Vidya Bhavan, Bombay, 1993)
\bibitem{majumdar5} R. C. Majumdar (General Editor), {\sl The History and
  Culture of the Indian People\/}, Vol. V, ({\sl The Struggle for
  Empire\/}), 4-th Edition, (Bharatiya Vidya Bhavan, Bombay, 1989)
\bibitem{saha} M. N. Saha (Chairman), {\sl Report of the Calendar Reform
  Committee\/}, (Government of India, New Delhi, 1955)
\bibitem{sengupta} P. C. Sengupta, {\sl Ancient Indian Chronology\/}
\pagebreak
\newpage
\bibitem{smart} W. H. Smart, {\sl Text Book on Spherical Astronomy\/},
 (University Press, Cambridge, 1931)
\bibitem{varaha1} Var\={a}hamihira, {\sl B\d{r}hat-sa\.{m}hit\={a}\/},
 English translation by Subrahmanya Sastri and M. Ramakrishna Bhat,
 2 Vols, (Bangalore, 1947)
\bibitem{varaha2} Var\={a}hamihira, {\sl Pa\~{n}ca Siddh\={a}ntik\={a}\/},
 Edited by G. Thibaut and S. Dvivedi, (Benares, 1889) 
\bibitem{winternitz} M.Winternitz, {\sl A Short History of Indian Literature\/},
 Vol. I, (University of Calcutta, Calcutta,1927)
\end{thebibliography}
\end{document}